\documentclass[twocolumn,preprintnumbers,amsmath,amssymb,showpacs]{revtex4}

\usepackage{graphicx}
\usepackage{dcolumn}
\usepackage{bm}

\newcommand{\stateFrequency}{{f}}
\newcommand{\probabilitySlab}[1]{\{\mathcal{P}_{k}\}^{#1}}
\newcommand{\piOne}{\pi_1}
\newcommand{\piMin}{\pi_{\rm min}}
\newcommand{\PiOne}{\Pi_1}
\newcommand{\xMin}{y}

\begin{document}
\title{Understanding the Frequency Distribution of Mechanically Stable 
Disk Packings}

\author{Guo-Jie Gao$^1$}
\author{Jerzy B{\l}awzdziewicz$^1$} 
\author{Corey S. O'Hern$^{1,2}$}
\affiliation{$^1$~Department of Mechanical Engineering, Yale University, 
New Haven, CT 06520-8284.\\
$^2$~Department of Physics, Yale University, New Haven, CT 06520-8120.\\
}
\date{\today}

\begin{abstract}
Relative frequencies of mechanically stable (MS) packings of
frictionless bidisperse disks are studied numerically in small
systems.  The packings are created by successively compressing or
decompressing a system of soft purely repulsive disks, followed by
energy minimization, until only infinitesimal particle overlaps
remain.  For systems of up to $14$ particles most of the MS packings
were generated.  We find that the packings are not equally probable as
has been assumed in recent thermodynamic descriptions of granular
systems.  Instead, the frequency distribution, averaged over each
packing-fraction interval $\Delta \phi$, grows exponentially with
increasing $\phi$.  Moreover, within each packing-fraction interval MS
packings occur with frequencies $\stateFrequency_k$ that differ by
many orders of magnitude.  Also, key features of the frequency
distribution do not change when we significantly alter the
packing-generation algorithm---for example frequent packings remain
frequent and rare ones remain rare.

These results indicate that the frequency distribution of MS packings
is strongly influenced by geometrical properties of the
multidimensional configuration space.  By adding thermal fluctuations
to a set of the MS packings, we were able to examine a number of local
features of configuration space near each packing.  We measured the
time required for a given packing to break to a distinct one, which
enabled us to estimate the energy barriers that separate one packing
from another.  We found a gross positive correlation between the
packing frequencies and the heights of the lowest energy barriers
$\epsilon_0$; however, there is significant scatter in the data.  We
also examined displacement fluctuations away from the MS packings to
assess the size and shape of the local basins near each packing.  The
displacement modes scale as $d_i \sim \epsilon_0^{\gamma_i}$ with
$\gamma_i$ ranging from approximately $0.6$ for the largest
eigenvalues to $1.0$ for the smallest ones.  These scalings suggest that
the packing frequencies are not determined by the local volume of
configuration space near each packing, which would require that the
dependence of $f_k$ on $\epsilon_0$ is much stronger than the
dependence we observe.  The scatter in our data implies that in
addition to $\epsilon_0$ there are also other, as yet undetermined
variables that influence the packing probabilities.
\end{abstract}

\maketitle

\section{Introduction}
\label{introduction}

Despite intense study over the past several decades, glassy and
amorphous materials are still poorly understood.  For example, a
fundamental explanation for the stupendous rise in the viscosity of
fragile glass-forming liquids as the temperature is lowered near the
glass transition is still lacking \cite{debenedetti}.  Also, the
response of glassy and amorphous systems to applied stress is 
difficult to predict because these systems display complex
spatio-temporal dynamics, such as shear bands \cite{fu,varnik},
strongly non-affine and cooperative motion \cite{maloney,tanguy}, and dynamical
heterogeneities \cite{ediger}.  Even basic questions concerning the
nature of stress and structural relaxation have not been adequately
addressed.  Important open questions include 1) what are the
characteristic rearrangement events that lead to stress and structural
relaxation, 2) how many particles are involved in such rearrangement
events, and 3) are these events correlated and over what length and
time scales?

An extremely useful concept for understanding the dynamical and
mechanical properties of glassy systems has been the potential energy
landscape (PEL) formalism \cite{stillinger}.  The PEL is the highly
multi-dimensional potential energy function that depends on all of the
configurational degrees of freedom of the system.  It has been shown
that the equation of state \cite{shell,nave} and dynamical quantities
\cite{keyes,heuer} such as the diffusion constant and viscosity
of supercooled liquids and glasses can be calculated in terms of
geometrical features of the PEL like local potential energy minima and
low-order saddle points. Related studies have also been performed on
hard spheres to understand the glass transition in these model
systems.  A significant focus of this research has been to explain
kinetic arrest in terms of decreasing free volume
\cite{salsburg,sastry2} and configurational entropy
\cite{speedy_chem_phys,speedy_biophys_chem} near the glass transition.

In this article we investigate particle packings and features of the
potential energy landscape in 2d bidisperse systems of frictionless
disks.  The disks interact via a finite-range continuous repulsive
potential.  We focus on mechanically stable {\sl (MS)} packings
with vanishing particle overlaps.  In these mechanically stable
packings any particle displacement results in an increase of the
potential energy, i.e., leads to overlap between particles.  Thus the
set of MS packings in our system is equivalent to the set of
collectively jammed states \cite{torquato} for hard disks.  However,
since we consider particles that interact via a continuous potential,
we can explore not only geometrical features of configuration space in
the neighborhood of a given collectively jammed state, but also
properties of the energy landscape near a given mechanically stable
packing.

An important feature of the present work (and our related earlier
study \cite{xu}) is that we focus on small systems containing $14$ or
fewer particles.  Related studies of small hard disk systems have been
carried out previously \cite{bs,bsv}, but these did not address
questions concerning the frequency with which MS packings occur, which
is the main focus of this work.  We confine our studies to small
systems for several reasons.  First, we believe that understanding
properties of small nearly jammed systems is crucial to developing a
theoretical explanation for slow stress and structural relaxation in
large glassy and amorphous systems.  Second, in small systems we are
able to generate nearly all of the mechanically stable disk packings,
which enables us to accurately measure the frequency with which
different MS packings occur.  Also, detailed results obtained for
small systems of different size can be extrapolated and used to infer
properties of glassy materials in the large-system limit.

Results from a number of recent studies emphasize that small
subsystems are important for understanding the slow dynamics displayed
in supercooled liquids and glasses.  For example, experiments on
colloidal glasses \cite{weeks1,weeks2} show that slow relaxation in
glassy materials occurs through local cage breaking events.  Caging
behavior has also been shown in a number of computer simulations of
hard and soft particles, for example in
Refs.~\cite{doliwa,thirumalai,miyagawa}.

Another indication that small subsystems play an important role in
determining the dynamics of glass-forming liquids can be found in the
results of recent numerical simulations of large binary hard-disk
systems in Ref.~\cite{donev}.  Using an appropriately defined
signature of the kinetic glass transition, these authors have shown
that a slowly quenched system falls out of equilibrium at packing
fractions $\phi$ in the range $0.77\alt\phi\alt0.8$ (the larger
packing fractions correspond to slower quench rates).  As revealed by
our recent study \cite{xu} (see also the results presented herein),
this is the packing-fraction range where the maximum of the density of
MS packings occurs for small systems of $N\agt12$ particles.  We
believe that this is not a coincidence.

The important role of small subsystems in dense hard-particle fluids
stems from the fact that small systems become geometrically jammed
(i.e., the system cannot rearrange) at packing fractions where
significant free volume is still available for the motion of particles
in their local cages.  In contrast, for macroscopic systems even
infinitesimal free volume per particle makes structural rearrangements
geometrically possible.

Such observations suggest that one should seek macroscopic
descriptions of glass-forming materials in terms of many coupled,
nearly jammed small subdomains.  At a packing fraction at which small
subdomains become geometrically jammed, rearrangements can occur only
if a domain is sufficiently uncompressed.  Large local density
fluctuations, however, occur infrequently, because of the low
compressibility of the surrounding domains at packing fractions above the
dynamic glass transition.  Hence, the structural relaxation time is
extremely large.  A picture similar to the one put forward by
Adam and Gibbs over $40$ years ago \cite{adam} can also be
constructed for systems of particles interacting via continuous
potentials.  In this case, relevant packing fractions can be defined
using a temperature-dependent effective particle diameter. 

A small subsystem of a macroscopic system corresponds to a subspace in
configuration space.  Thus, an approach that describes the dynamics of
glass-forming liquids in terms of a set of small coupled subsystems
should help to reconcile the local picture of glassy materials (e.g.,
trap models \cite{bouchaud}) with the global PEL view.  However,
before predictive theories can be constructed based on the ideas
outlined above, one needs to obtain a more complete understanding of
the behavior of small nearly jammed systems.

An analysis of MS packings in small systems is relevant not only for
understanding the slow dynamics of glass-forming liquids, but also for
explaining the meaning of random-close packing in macroscopic athermal
amorphous materials.  We now turn to this aspect of the problem.

In MS packings composed of touching (but not overlapping) disks, the
number of degrees of freedom is less than or equal to the number of
constraints.  Thus, MS packings, or collectively jammed states, can be
represented by single points in configuration space.  In small systems
we are able to generate nearly all mechanically stable disk packings.
Thus, we can investigate, separately, the number of distinct packings that
exist in a given interval of packing fraction $\phi$ (i.e., the
density of MS packings) and the probability with which these packings
occur for a given generation protocol \cite{xu}.

An analysis of the protocol-independent density of MS packings and
protocol-dependent probabilities for each distinct packing allows us
to address the question of why random close packing seems to be a well
defined quantity (i.e., many distinct generation protocols give
similar values for it, for example in
Refs.~\cite{ohern_short,ohern_long,silbert,makse}), but it is also a
poorly defined concept because other protocols, for example, those
that allow slow thermal quench rates, yield a continuous range of
packing fractions at which jammed states occur \cite{torquato2}.

In our opinion, questions concerning random close packing have not yet
been fully resolved.  In particular, although the maximally random
jammed (MRJ) state as described in Ref.~\cite{torquato2} is a useful
and important concept, we believe that it is not the final answer.
For example, MRJ involves an arbitrary choice of how to characterize
order in the system and does not address the question of why a wide
class of generation protocols yield similar values for random-close
packing.

In our recent paper \cite{xu} we argued that the random close packing
density corresponds to the position of the peak in the density of MS
packings $\rho(\phi)$, which becomes a $\delta$-function in the
infinite system-size limit.  As opposed to the protocol-dependent
probability density $P(\phi)$ for obtaining a packing at a given
$\phi$, the density of MS packings is a {\it protocol-independent\/}
quantity.  Unless a given protocol is strongly biased toward states
in the tail of the distribution $\rho(\phi)$, for example those
protocols that involve significant thermalization, a well-defined
random close packed density is obtained in the large system limit.
This explains why many quite distinct protocols consistently yield
similar values for the random close packed density $\phi_{\rm rcp}$.

\begin{figure}
\scalebox{0.5}{\includegraphics{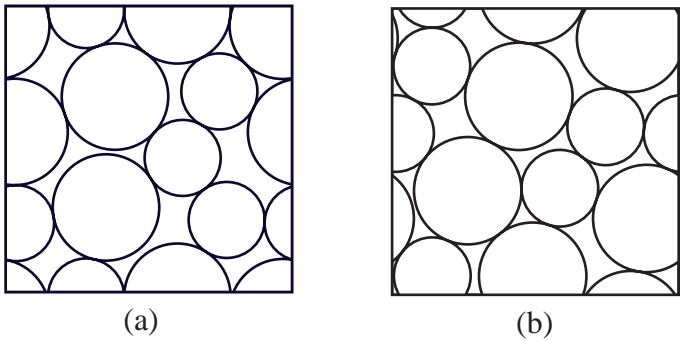}}%
\vspace{-0.18in}
\caption{\label{fig:compare} Snapshot of (a) a rare mechanically
stable packing at $\phi \approx 0.82$ and (b) another MS packing $\phi
\approx 0.83$ that is $10^6$ times more frequent.}
\vspace{-0.22in}
\end{figure}

This explanation, while quite plausible, leaves a number of important
issues unresolved.  For example, we find that individual MS packings
occur with extremely different frequencies even for a typical
fast-quench protocol \cite{xu}.  Two MS packings at approximately the
same packing fraction $\phi$ but with drastically different
frequencies are shown in Fig.\ \ref{fig:compare}.  There are no
striking structural differences between these two MS packings, yet, the
packing shown in Fig.\ \ref{fig:compare}(a) is $10^6$ times less
frequent than that in Fig.\ \ref{fig:compare}(b).
Moreover, we find that there are many more infrequent MS packings than
frequent ones.  

An important issue is what determines whether a particular MS packing
is frequent or rare.  Although it is clear that the particular
protocol chosen to generate the MS packings plays a role in
determining the frequency distribution \cite{torquato2}, we argue that
geometrical features of the PEL also strongly influence the frequency
distribution.  First, our protocols, which involve a sequence of
compression and decompression steps followed by energy minimization do
not target any particular MS packings.  Yet, the probabilities of different
MS packings vary by many orders of magnitude.  Moreover, as we will show
below, even when we significantly alter the protocol used to generate
the MS packings, the most frequent packings remain frequent and the
rare packings remain rare.

The question of what gives rise to the extremely diverse probabilities
of different MS packings is not only important for the analysis of
random close packing, but also for developing statistical
descriptions of granular materials and other athermal systems.  For
example, it has been assumed in Edwards' entropy descriptions of
granular media \cite{mehta,makse2} that different jammed packings
occur with approximately equal weights.  Similar assumptions are also
usually employed in PEL theories of glassy materials
\cite{debenedetti2}.  However, these assumptions should be reexamined,
since analogous geometric features are likely to control the frequency
distributions of relevant states in these systems as in our studies of
the frequency of MS packings.  Understanding the reason for the
enormous probability differences and the comparative roles of the frequent and
infrequent states are thus important unresolved problems.

This paper is organized as follows.  In Sec.~\ref{Mechanically stable
packings in bidisperse systems of disks}, we describe the model and
protocol we use to generate the MS packings and how we classify and count
each distinct packing.  In Sec.~\ref{Packing frequency distribution},
we review our results for the density of MS packings and their
frequency distribution in small 2d bidisperse systems.  In
Sec.~\ref{energy_landscape}, we investigate the correlation between
local geometric features of configuration space near each MS packing
and the MS packing frequencies by adding thermal
fluctuations to the packings.  We look for structural properties
that distinguish between frequent and rare MS packings in
Sec.~\ref{structure}.  In Sec.~\ref{Hidden variable} we introduce a simple
phenomenological model to explain the dramatic variation in MS packing
frequencies.  We conclude and briefly discuss future research
directions in Sec.~\ref{conclusions}.

\section{Generation and classification of mechanically stable
disk packings}
\label{Mechanically stable packings in bidisperse systems of disks}

\subsection{Model}
\label{The system}

We study the mechanical and statistical properties of MS packings in
two-dimensional bidisperse systems of $N$ frictionless disks
interacting via the finite-range, pairwise additive, purely repulsive
spring potential of the form
\begin{equation}
\label{spring potential}
V(r_{ij})
=\frac{\epsilon}{2}(1-r_{ij}/\sigma_{ij})^2\Theta(\sigma_{ij}/r_{ij}-1).
\end{equation}
Here $\epsilon$ is the characteristic energy scale, $r_{ij}$ is the
separation between particles $i$ and $j$, $\sigma_{ij}=\left(\sigma_i
+ \sigma_j \right) /2$ is their average diameter, and $\Theta(x)$ is
the Heaviside step function.  The particles are in a square unit cell
with periodic boundary conditions.  The mass $m$ of all particles is
assumed to be equal.

We consider $50$-$50$ binary mixtures of large and small particles
with diameter ratio $1.4$. We study bidisperse mixtures because the
presence of particles with different sizes inhibits crystallization
(provided that the size ratio is incommensurate with a binary crystal
structure).  In contrast, monodisperse disk packings crystallize
readily \cite{berryman,connelly}, so these systems have entirely
different properties than glass-forming liquids that are the subject of
our investigations.  Bidisperse mixtures of disks with diameter
ratio $1.4$ have been used in several previous studies
\cite{speedy,ohern_short,ohern_long,xu,donev}. Such mixtures thus constitute a
convenient reference system for investigations of fundamental
properties of amorphous and glassy materials.

In systems with finite-range purely repulsive interparticle potentials there
are two general classes of potential energy minima.  First, the system can
possess a connected network of particle overlaps that spans the system; the
sum of forces on each particle in such networks is zero.  These configurations
have positive total potential energy, and each displacement of the particles
in the network results in an energy increase.  For the second type there are no
particle overlaps, all forces are zero, the total potential energy vanishes,
and the particles can be moved without creating an overlap.

Our focus here is on states that are on the border between these two
general classes: we assume that the system is at an energy minimum
with infinitesimal particle overlaps (thus, the total energy is also
infinitesimal).  Since these states are assumed to be in stable
mechanical equilibrium and possess vanishingly small overlaps, we term these
states MS packings. 

If all particles of a MS packing participate in the system spanning
force network, no particle displacement is possible without creating
an overlap.  
Occasionally, a small number of particles in a MS packing (no more
than three for the small systems considered herein) do not participate
in the force network.  States with such free particles (rattlers) have
to be treated with care in the packing generation
procedures. However, we do not find that the properties of MS packings
with rattlers deviate significantly from those of MS packings without
rattlers.

\subsection{Packing generation protocols}
\label{Generation protocols}

\subsubsection{Simulation algorithms}

We use here a class of packing-generation protocols that involve successive
compression or decompression steps followed by energy minimization
\cite{makse,xu}.  The system is decompressed (or, equivalently, the particle
diameters are decreased) when the energy of the system at a
local minimum is nonzero; otherwise, the system is compressed.  The increment
by which the particle packing fraction is changed at each compression or
decompression step is gradually decreased.  After a sufficiently large number
of steps, a MS packing with infinitesimal overlaps is thus obtained.

For each independent trial, we begin the process by choosing random
initial positions for the particles at packing fraction $\phi_0 =
0.60$ (which is well below the minimum packing fraction at which MS
packings occur in 2d).  We then allow the system to relax, and perform
a sequence of compression/decompression and relaxation steps.  We can
repeat this process for many independent initial conditions, generate
a large number of MS packings, and measure their respective probabilities
for a given protocol.

In the present study, we use two energy-minimization methods: (a)
conjugate-gradient (CG) minimization algorithm or (b) molecular dynamics
(MD) with dissipation proportional to local velocity differences.  The
conjugate-gradient method is a numerical scheme that begins at a given
point in configuration space and moves the system to the nearest local
potential energy minimum without traversing any energy barriers
\cite{numrec}.  In contrast, molecular dynamics with finite damping is
not guaranteed to find the nearest local potential energy minimum
since kinetic energy is removed from the system at a finite rate.  The
system can thus surmount a sufficiently low energy barrier.  In the
molecular dynamics method, each particle $i$ obeys Newton's equations
of motion
\begin{equation} \label{newton1}
m {\vec{a}}_i = 
\sum_{j\not=i}
   \Theta(\sigma_{ij}/r_{ij}-1) \left[\frac{\epsilon}{\sigma_{ij}}
\left(1-\frac{r_{ij}}{\sigma_{ij}} \right) - b\vec{v}_{ij} \cdot
{\hat{r}}_{ij} \right] {\hat{r}}_{ij},
\end{equation}
where $\vec{a}_i$ is the acceleration of particle $i$, $\vec{v}_{ij}$
is the relative velocity of particles $i$ and $j$, ${\hat{r}}_{ij}$ is
the unit vector connecting the centers of these particles, and $b$ is
the damping coefficient.  In the present study, we chose the
dimensionless damping coefficient $\bar b=\sigma b/\sqrt{m\epsilon} =
0.5$, but this will be varied in subsequent studies.  
In the infinite-dissipation limit $b\to \infty$ the potential energy
cannot increase during a molecular-dynamics relaxation, and thus
the molecular-dynamics and conjugate-gradient methods should
give very similar results.  We note, however, that even in this limit
the two methods are not equivalent because there may be more than one
energy minimum accessible from a given point in configuration space
without traversing an energy barrier.  In our previous studies
\cite{ohern_short,ohern_long,xu}, we used only the conjugate-gradient
minimization algorithm.

\subsubsection{Implementation details}

In specific implementations of our MS-packing generation protocols, one needs
to use appropriate numerical criteria for stopping the energy minimization
process either in a configuration with overlapping particles forming a force
network or in a state with no particle overlaps.  For the the conjugate
gradient method, we terminate the minimization process when either of the
following two conditions on the potential energy per particle $V$ is
satisfied: (a) two successive conjugate gradient steps $n$ and $n+1$ yield
nearly the same energy value, $(V_{n+1} - V_{n})/V_n < \delta = 10^{-16}$; or
(b) the potential energy per particle at the current step is extremely small,
$V_n < V_{\rm min} = 10^{-16}$ (where $V$ is normalized by the energy-scale
parameter $\epsilon$).  Since the potential energy oscillates in time in the
molecular dynamics method, condition (a) is replaced by the requirement that
the relative potential-energy fluctuations satisfy the inequality $\langle (V
- \langle V \rangle)^2 \rangle^{1/2}/\langle V \rangle < \delta$.

Following the energy minimization, we determine whether the system should be
compressed or expanded.  If, $V \le V_{\rm min}$, the system is below the
jamming threshold, and thus it is compressed by $\Delta \phi$.  If, on the
other hand, $V > V_{\rm max} = 2V_{\rm min}$, the system is decompressed by
$\Delta \phi$.  For the first compression or decompression step we use the
packing-fraction increment $\Delta\phi = 10^{-4}$.  Each time the procedure
switches from expansion to contraction or vice versa, $\Delta \phi$ is reduced
by a factor of $2$.  

For the molecular-dynamics procedure to be efficient, rattler particles with
no contacts must be treated with care.  When the system is near the jamming
threshold, we set the velocities of rattlers to zero; we also shift the
center-of-mass velocity of the remaining non-rattler particles to zero to
assure momentum conservation.  This modification of the energy-minimization
procedure allows our stopping criteria to be implemented without change even
when rattlers are present.  Otherwise, the kinetic energy of a rattler decays
too slowly, and it is extremely difficult for the system to reach the 
threshold $V_{\rm min}$.

The MS packing generation process terminates when $V_{\rm
min}<V<V_{\rm max}$ after energy minimization.  In the final state,
the system is thus in mechanical equilibrium with extremely small
overlaps in the range $10^{-9} < 1- r_{ij}/\sigma_{ij} < 10^{-8}$.
We verify the stability of each final equilibrium configuration by
calculating the dynamical or Hessian matrix (i.e., the matrix of second
derivatives of the total potential energy with respect to the particle
positions) \cite{alexander}.  In a small percentage of cases we find
that the dynamical matrix has extra zero eigenvalues that do not
correspond to rattlers, which indicates that the system is at a saddle
point rather than at an energy minimum.  Such unstable packings are
not considered.  A more detailed discussion of the dynamical (Hessian)
matrix is given in Sec.\ \ref{classify}.

Our procedure for finding mechanically stable disk packings allows us
to determine the jamming threshold in $\phi$ to within $10^{-8}$.  The
procedure is similar to those implemented recently for static granular
packings with and without friction \cite{silbert,makse}.  Our results,
however, have much greater precision.  High accuracy is important in
our packing-enumeration studies, because some of the distinct MS
packings have nearly the same packing fraction.

\begin{figure}
\scalebox{0.48}{\includegraphics{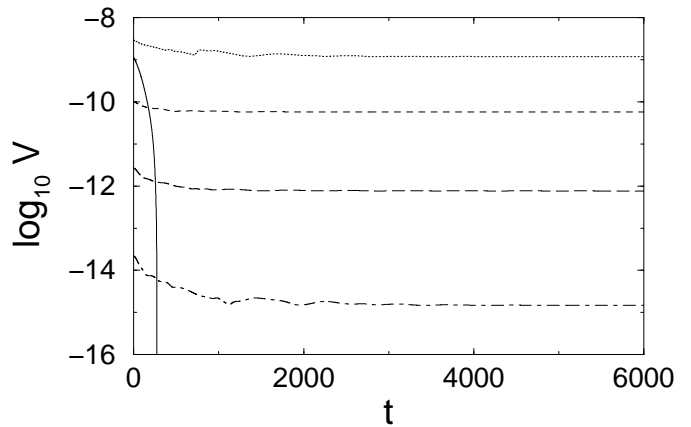}}%
\vspace{-0.18in}
\caption{\label{fig:V_t} Potential energy per particle $V$ as a
function of time $t$ at several $\phi$ during the process of creating
a mechanically stable packing of $10$ particles at $\phi_c \approx
0.8244438$.  The curves correspond to $\phi \approx 0.8245$ (dotted),
$0.82443$ (dashed), $0.82444$ (long-dashed), $0.824443$ (dot-dashed),
and $0.8230$ (solid).  MD energy minimization was employed.  Note that
for $\phi > \phi_c$ energy minimization terminates at $V>V_{\rm
min}=10^{-16}$, while for $\phi < \phi_c$ minimization terminates when
$V=V_{\rm min}$.  The spring timescale $t_s = \sigma
\sqrt{m/\epsilon}$, where $\sigma$ is the small particle diameter, was
chosen as the normalization for time.}
\vspace{-0.22in}
\end{figure}

To illustrate typical behavior of the system during the MS-packing
generation process in the version with molecular-dynamics energy
minimization procedure we show, in Fig.~\ref{fig:V_t}, the evolution
of the potential energy per particle $V$ for several consecutive
values of $\phi$.  Note that the potential energy is not
monotonic in time because of finite particle inertia.

\subsection{Identification of Distinct MS Packings}
\label{classify}

We consider two MS packings to be identical if they have the same network of
contacts (i.e., their networks can be mapped onto each other by translation,
rotation, or by permutation of particles of the same size).  An analysis of
the topology of the network of contacts is, however, a fairly complex task.
Thus, in practice it is convenient to use some alternative but equivalent
criteria.

A test based on the packing fraction alone is inadequate, because some
packings with distinct topological networks have the same packing fraction
(within the numerical noise).  We find, however, that a test based on
a comparison of the spectra of the dynamical matrix is sufficient.

For a pairwise additive, rotationally invariant potential (\ref{spring
potential}) the dynamical matrix (Hessian) is given by the expressions
\cite{tanguy}
\begin{equation}
\label{offdiagonal}
M_{i\alpha,j\beta}
  =-\frac{t_{ij}}{r_{ij}}(\delta_{\alpha \beta}
    -{\hat r}_{ij\alpha}{\hat r}_{ij\beta})
    -c_{ij} {\hat r}_{ij\alpha} {\hat r}_{ij\beta},
                 \quad i\not=j,
\end{equation}
and
\begin{equation}
\label{diagonal}
M_{i\alpha,i\beta} = - \sum_{j\not=i} M_{i\alpha,j\beta},
\end{equation}
where $t_{ij} = \partial V/\partial r_{ij}$ and $c_{ij} = \partial^2 V/
\partial r_{ij}^2$.  In the above relations the indices $i$ and $j$ refer to
the particles, and $\alpha,\beta=x,y$ represent the Cartesian coordinates.
For a system with $N_f$ rattlers and $N'=N-N_f$ particles forming a
connected network the indices $i$ and $j$ range from $1$ to $N'$, because the
rattlers do not contribute to the potential energy.  

The dynamical matrix is symmetric, and it has $dN'$ rows and columns, where
$d=2$ is the spatial dimension.  Thus it has $dN'$ real eigenvalues $\{ m_i
\}$, $d$ of which are zero due to translational invariance of the system.  In
a mechanically stable disk packing, no set of particle displacements is
possible without creating an overlapping configuration; therefore the
dynamical matrix has exactly $d(N'-1)$ nonzero eigenvalues.  In our
simulations we use the criterion $|m_i| > m_{\rm min}$ for nonzero
eigenvalues, where $m_{\rm min} = 10^{-6}$ is the noise threshold for our
eigenvalue calculations.

In our numerical simulations we distinguish distinct mechanically stable disk
packings by the lists of eigenvalues of their dynamical matrices.  We assume
that two MS packings are the same if and only if they have the same list of
eigenvalues.  (The eigenvalues are considered to be equal if they differ
by less than the noise threshold $m_{\rm min}$.)  To verify our method we
have also compared the topology of the network of particle contacts in
different packings, and we have found that these two methods of identifying
distinct mechanically stable packings agree.

As noted above, it is in general not true that each distinct MS packing
possesses a unique packing fraction $\phi$.  However, we find that for small
systems only at most a few percent of distinct MS packings share the same
packing fraction.  Thus, in the following we will associate a unique $\phi$
with each MS packing to simplify the discussion.  Also, approximately $10\%$
of the distinct MS packings contain at least one rattler particle.  In these
configurations, we ignore the translational degeneracy of the rattlers---two
configurations that have the same contact networks of non-rattler particles
are treated as the same.  We do not find that the properties of the MS
packings with rattlers deviate significantly from those of the MS packings
without rattlers.

\begin{table}
\caption{\label{tab:table1} Number of distinct MS packings $n_s^{\rm
MD}$ and $n_s^{\rm CG}$ obtained and number of trials $n_t^{\rm MD}$
and $n_t^{\rm CG}$ performed using the MD and CG energy minimization
methods and estimated total number of distinct MS packings $n_s^{\rm
tot}$ calculated using an extrapolation of the CG results for several
system sizes $N$. }
\begin{ruledtabular}
\begin{tabular}{rrrrcc}
$N$  & $n_s^{\rm MD}$ & $n_s^{\rm CG}$ & $n_s^{\rm tot}$ & $n_t^{\rm MD}$ 
& $n_t^{\rm CG}$ \\
\tableline
$4$ & $3$ & $3$ & $3$ & $10^5$ & $10^5$ \\ 
$6$ &  $20$ & $20$ & $20$ & $1.5 \times 10^6$ & $10^6$ \\
$8$ &  $155$ & $165$ & $165$ & $14 \times 10^6$ & $10^6$ \\
$10$ & $1247$ & $1618$ & $1618$ & $30 \times 10^6$ & $30 \times 10^6$ \\
$12$ & --- & $23460$ & $26100$ & --- & $28 \times 10^6$ \\
$14$ & --- & $248900$  & $371500$ & --- & $48 \times 10^6$ \\
\end{tabular}
\end{ruledtabular}
\end{table}

\section{Probability distribution of MS packings}
\label{Packing frequency distribution}

\subsection{Total number of distinct MS packings}

We have applied our algorithm for finding mechanically stable disk packings
using a large number of independent trials with different starting
configurations for systems with up to $N=256$ particles.  Here we
focus, however, only on small systems with $N\le14$, because for these values
of $N$ we were able to generate a significant fraction of all distinct MS
packings.  Both the conjugate gradient and molecular dynamics energy
minimization techniques have been employed to determine the
dependence of the results on the packing-generation protocol.

In Table \ref{tab:table1} we list the numbers of distinct MS packings
$n_s^{\rm MD}$ and $n_s^{\rm CG}$ obtained using the MD and CG methods
and the corresponding numbers of trials performed $n_t^{\rm MD}$ and
$n_t^{\rm CG}$.  Our estimate for the the total number of MS packings
that exist $n_s^{\rm tot}$ for each system size is also given.

The total number of distinct MS packings has been estimated by
extrapolating the relation between $n_s^{\rm CG}$ and $n_t^{\rm CG}$
using the approach proposed in \cite{xu}.  For $N\le10$ the number of
distinct packings $n_s^{\rm CG}$ generated by the CG method for the
given number of trials $n_t^{\rm CG}$ saturates, which indicates that
nearly all MS packings have been obtained for these system sizes
\cite{rare}.  For $N=12$ and $14$ the CG method has yielded about 90\,\%
and 70\,\% of the total MS packings, respectively.

For the two smallest systems studied, $N=4$ and $6$, the sets of MS
packings generated by the CG and MD methods are identical.  For larger
system sizes, the CG method finds more packings than the MD method for a
fixed number of trials because a large fraction of MS packings become
extremely rare when using the MD method. (This interesting feature of
the MD packing-generation algorithm will be discussed in more detail
below.)  

The results in Table \ref{tab:table1} indicate that the number of
distinct MS packings grows exponentially with increasing system size.
In addition, the number of trials needed to find all MS packings (or
to find a large fraction of them) exceeds by orders of magnitude the
number of distinct MS packings themselves.  For example, with as many
as $3\times10^7$ MD trials, we have found only about $1250$ packings out
of approximately $1600$. The CG results show similar behavior, but
rare packings are not as rare as for MD.  These results indicate that
the frequencies with which MS packings occur can vary by many orders
of magnitude.  In this article we seek to understand the source and
the significance of this property.

\subsection{Frequency distribution and density of MS packings}
\label{The frequency distributions - subsection}

\subsubsection{Protocol-dependent and protocol-independent 
quantities}

\begin{figure}
\scalebox{0.4}{\includegraphics{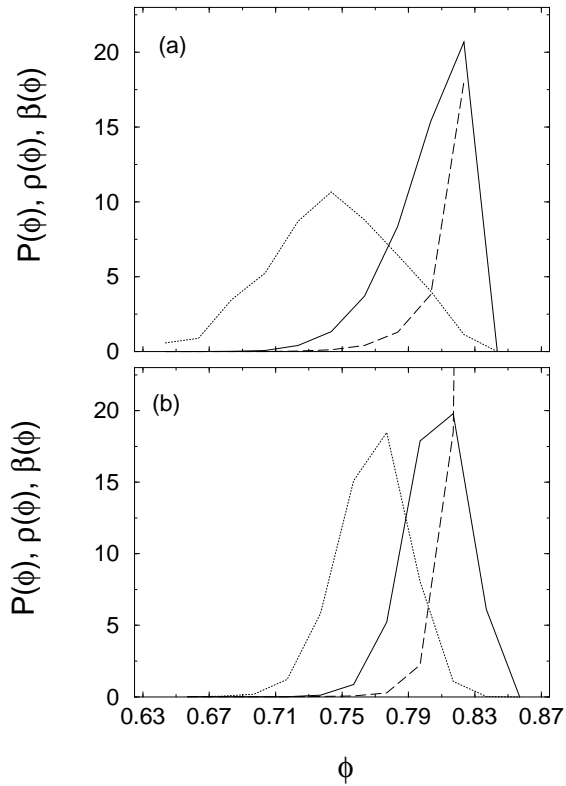}}%
\vspace{-0.18in}
\caption{\label{fig:decomposition} The probability density $P(\phi)$
(solid), density of MS packings $\rho(\phi)$ (dotted), and the
frequency distribution $\beta(\phi)$ (long-dashed) averaged over bins
with width $\Delta \phi = 0.02$ for (a) $10$- and (b) $14$-particle
systems using the CG energy minimization method.  Note that the 
results for $N=14$ are approximate since we have not found all of 
the MS packings for this system size.  
}
\vspace{-0.22in}
\end{figure}


Nearly complete enumeration of the MS packings for small systems
allows us to characterize the distribution of MS packings in much more
detail than is possible for a system with large $N$.  For sufficiently
small systems, each distinct MS packing can be generated multiple
times; thus for each distinct packing (indexed by $k$), we can
determine the protocol-dependent probability $\stateFrequency_k$ for
which it occurs.  Next, for a given packing-fraction interval
$\Delta\phi$ we can evaluate the probability
\begin{equation}
\label{probability}
P(\phi)\Delta\phi={n_P(\Delta\phi)}/{n_t}
\end{equation}
of generating a state in the interval $\Delta\phi$ for a specific
protocol.  We can also define the protocol-independent number of distinct
MS packings 
\begin{equation}
\label{densit_of_states}
\rho(\phi)\Delta\phi={n_s(\Delta\phi)}/{n_s}
\end{equation} 
that exist in this interval.  In Eq.\ \ref{probability},
$n_P(\Delta\phi)$ denotes the number of trials that produce MS
packings in the interval of interest and $n_t$ is the total number of
trials.  The corresponding quantities in Eq.\ \ref{densit_of_states}
are the number of distinct MS packings $n_s(\Delta\phi)$ that exist in
the given packing-fraction interval and the total number of distinct
MS packings $n_s$ in the system.  We note that both the probability
density $P(\phi)$ and the density of MS packings $\rho(\phi)$ are
normalized to unity.

In addition to the quantities defined by Eqs. \ref{probability} and
\ref{densit_of_states}, we also consider the average frequency
distribution
\begin{equation}
\label{frequency beta}
\beta(\phi)=P(\phi)/\rho(\phi)
\end{equation}
for MS packings in the interval $\Delta\phi$ near the packing fraction
$\phi$.  The continuous frequency distribution \ref{frequency beta}
and the discrete probabilities $\stateFrequency_k$ for obtaining the
$k$th packing satisfy the relation
\begin{equation}
\label{continuous beta in terms of discrete beta average}
\beta(\phi)=n_s\langle\stateFrequency_k\rangle,
\end{equation}
where 
\begin{equation}
\label{discrete beta average}
\langle\stateFrequency_k\rangle
  =
   \sum_{\phi_k\in\Delta\phi}\stateFrequency_k/n_s(\Delta\phi)        
\end{equation}
is the average value of the probability $\stateFrequency_k$ in the interval
$\Delta\phi$.

\begin{figure}
\scalebox{0.4}{\includegraphics{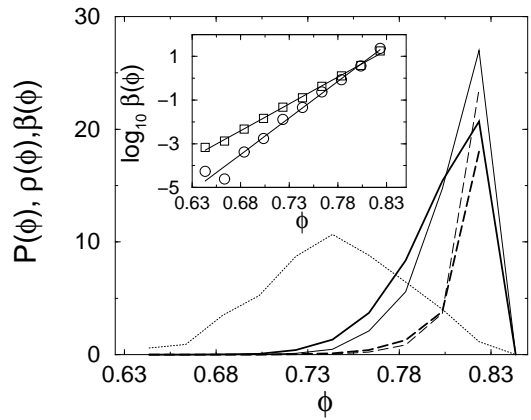}}
\caption{\label{fig: CG and MD frequencies} A comparison of the total
probability distribution $P(\phi)$ (solid) and frequency distribution
$\beta(\phi)$ (long-dashed) averaged over bins with width $\Delta \phi
= 0.02$ for a $10$-particle system using the CG (thick lines) and MD
(thin lines) methods.  The density of MS packings (dotted) is also
shown. The MD results only include approximately $75\%$ of the total MS
packings.  The inset shows $\beta(\phi)$ using the CG (squares) and MD
(circles) methods for the same system on a logarithmic scale.  The
slopes of the lines are approximately $25$ and $33$.}
\vspace{-0.22in}
\end{figure}


\subsubsection{Relation between density $\rho(\phi)$ of MS packings and 
random-close packing}

The decomposition of the probability density \ref{probability} into
the density of MS packings \ref{densit_of_states} and the average
frequency \ref{frequency beta} was studied previously in our recent article
\cite{xu} for the CG packing-generation protocol.  Sample results of
our simulations are shown in Fig.\ \ref{fig:decomposition}, where we
plot the functions $P(\phi)$, $\rho(\phi)$, and $\beta(\phi)$
for $N=10$ and $14$.

Two key features should be noted from this figure.  First, the
peaks of the density of MS packings $\rho(\phi)$ and probability distribution
$P(\phi)$ become narrower as the system size grows.  Second, the separation
between the peaks decreases with increasing $N$.  We observe that the distance
between the maxima of $\rho(\phi)$ and $P(\phi)$ is of the order of the peak
width both for $N=10$ and $14$.

This behavior suggests that the random-close packed density $\phi_{\rm
rcp}$ can be re-defined as the position of the peak in the density of
MS packings as we proposed in Ref.\ \cite{xu}.  The fluctuations in
the packing fraction of large MS packings can be described by a
superposition of local fluctuations in a large number of subsystems.
Therefore, the width of the peak in $\rho(\phi)$ should scale
approximately as $N^{-1/2}$ by the central-limit theorem, which was
confirmed in Ref.~\cite{ohern_long}.  The position of the peak in
$\rho(\phi)$ will coincide with the peak position of the probability
distribution $P(\phi)=\beta(\phi)\rho(\phi)$ in the large system
limit, unless the frequency $\beta(\phi)$ varies so rapidly with
$\phi$ that it elevates the tail of the distribution $\rho(\phi)$.

According to this definition, random close packing is a {\it
protocol-independent} quantity.  Our picture is thus consistent with
the observation that for a large class of protocols essentially the
same random close packed density is obtained for macroscopic systems.
This picture is also supported by the results shown in Fig.\ \ref{fig:
CG and MD frequencies}, where the probability density $P(\phi)$ is
depicted for the CG and MD protocols.  The results indicate that the
peaks in $P(\phi)$ nearly coincide for the two protocols, in spite of
a noticeable difference between the corresponding frequency
distributions $\beta(\phi)$ for the MD and CG methods, especially at the lower
end of the range in packing fraction.

\begin{figure}
\scalebox{0.5}{\includegraphics{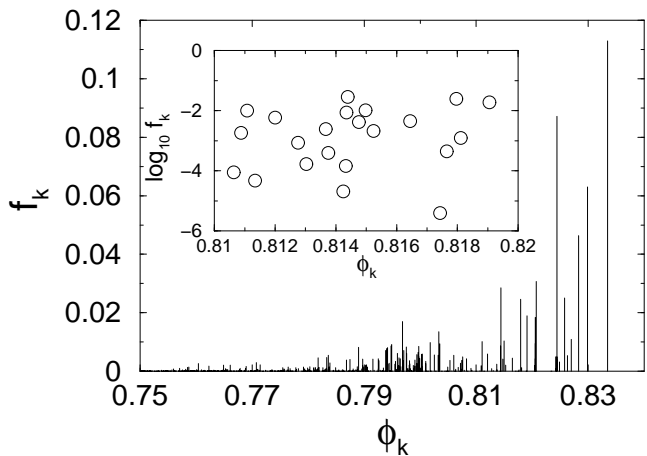}}%
\vspace{-0.2in}
\caption{\label{fig:TD} The discrete probability $f_k$ to obtain a MS
packing at a given packing fraction $\phi_k$ for $N=10$ particles using the
MD energy minimization method. The inset shows a magnified view of the
probability on a logarithmic scale over a narrow region of packing fraction
between $0.81$ and $0.82$.  Even within this narrow region, the
probability varies by more than five orders of magnitude.}
\vspace{-0.22in}
\end{figure}


\begin{figure}
\scalebox{0.4}{\includegraphics{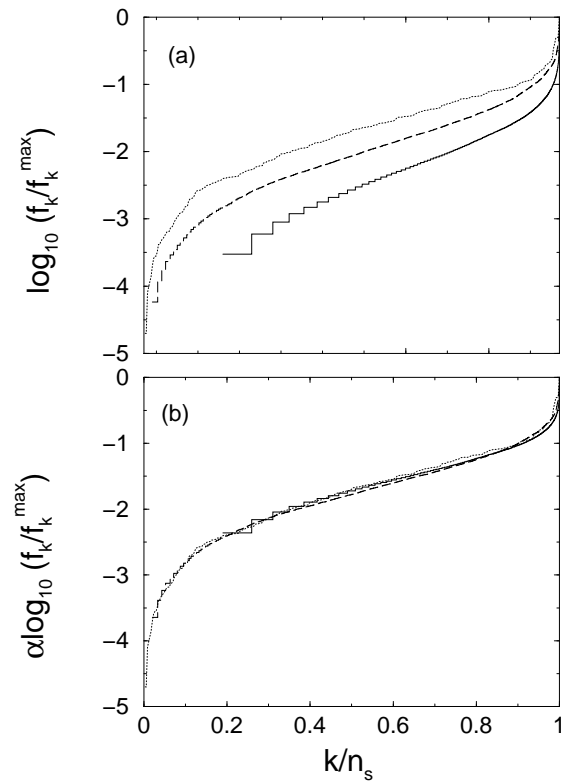}}%
\vspace{-0.05in}
\caption{\label{sorted probabilities for different N} (a) The discrete
probabilities $f_k/f_k^{\rm max}$ for MS packings obtained using the
CG method within a narrow interval $\Delta \phi$ near the peak in
$\rho(\phi)$ normalized by the maximal probability in the interval
$f_k^{\rm max}$, sorted in increasing order, and plotted on a
logarithmic scale for $N=10$ (dotted), $12$ (long-dashed), and $14$
(solid) particles.  The interval $\Delta \phi = 0.02$ for $N=10$ and
$12$ and $0.004$ for $N=14$.  The index $k$ on the horizontal axis is
normalized by the total number of distinct MS packings $n_s$ that
exist within $\Delta \phi$. (b) The probabilities $\log_{10} \left(
f_k/f_k^{\rm max} \right)$ scaled by an $O(1)$ constant $\alpha$.
$\alpha$ and $n_s$ were chosen to yield the best collapse of the
scaled curves for $N=12$ and $14$ with the unscaled curve for $N=10$.}
\vspace{-0.05in}
\end{figure}


The above-described scenario seems quite plausible, but there are
still a number of puzzling findings that need to be addressed before
amorphous jammed packings are fully understood.  First, we find that
even for the class of algorithms considered here the frequency
$\beta(\phi)$ is a rapidly varying, exponential function of $\phi$.
Such exponential behavior of $\beta(\phi)$ is clearly seen in the
inset of Fig.\ \ref{fig: CG and MD frequencies} for the $10$-particle
system.  Similar results were also obtained for larger $N$.

The results shown in Fig.\ \ref{fig: CG and MD frequencies} indicate
that the exponential variation of the frequency distribution
$\beta(\phi)$ is quite rapid---we find that this quantity changes by
several order of magnitude in the relatively narrow range of packing
fractions where a significant number of MS packings exists.  Moreover,
the exponent is an increasing function of $N$, as revealed by
simulations presented in Ref.~\cite{xu}.  Under the assumption that
the exponential behavior also holds for large systems, the peak of the
probability $P(\phi)$ is controlled by the peak of the density
$\rho(\phi)$ of MS packings, provided that $\log\beta/N\to0$ for
$N\to\infty$.  With current computational resources we are not yet
able to assess the validity of this condition.  The source of the
rapid variation of $\beta(\phi)$ and the system-size dependence
of this variation are thus important unresolved issues.

Our simulations also reveal another striking result.  We find that the
discrete probabilities of MS packings can differ by many orders of
magnitude even when they possess nearly the same value of $\phi$.  This
behavior will now be discussed.

\subsubsection{Discrete probabilities of  distinct MS packings}
\label{Numerical results for frequency distributions}

The discrete probabilities $\stateFrequency_k$ of distinct MS packings
in the range of packing fractions within the peak of the probability
density $P(\phi)$ are depicted in Fig.\ \ref{fig:TD} for the
$10$-particle system.  The results are shown for the MD
packing-generation protocol; the corresponding distribution 
for the CG protocol is similar.

Consistent with our findings for the continuous frequency distribution
$\beta(\phi)$ the probabilities $\stateFrequency_k$ are, on average,
significantly larger at larger values of $\phi$.  The discrete
probabilities $\stateFrequency_k$, however, also vary dramatically
from one MS packing to another.  They can differ by more than five orders of
magnitude even in a narrow interval of $\phi$, as illustrated in the inset
of Fig.\ \ref{fig:TD}.  These large local probability fluctuations occur
over the entire range of $\phi$.

\begin{figure}
\scalebox{0.4}{\includegraphics{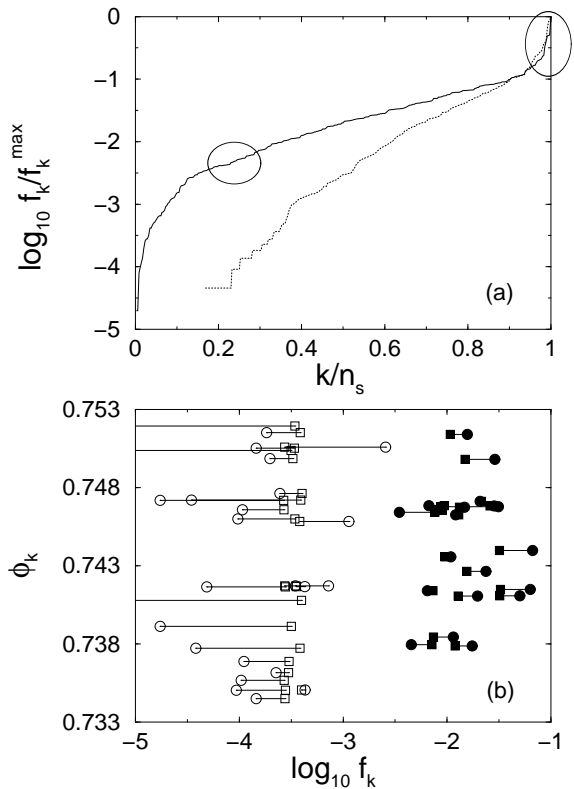}}
\caption{\label{sorted probabilities for different protocols} 
(a) A comparison of the sorted discrete probabilities $f_k$ normalized
by $f_k^{\rm max}$ for the CG (solid line) and MD (dotted line)
methods for an interval in packing fraction $\Delta \phi$ near the
peak in $\rho(\phi)$ for $N=10$.  (b) The discrete probabilities $f_k$
for the $20$ most frequent MS packings depicted in (a) from the CG
packing-generation method (filled squares) are compared to the
probabilities for the same packings obtained from the MD method
(filled circles).  A similar comparison of the probabilities for $25$
less frequent packings obtained from the CG (open squares) and MD
(open circles) methods are also shown.  The circled regions in (a)
identify the MS packings that were compared in (b).
}
\end{figure}


It is clear from these results that MS packings do not occur with
approximately equal probabilities as has been assumed in the Edwards'
entropy descriptions of powders and granular media
\cite{mehta,makse2}.  The large variation of the probabilities is a
rather puzzling result because our algorithms do not target specific
packings.  This suggests that the probabilities are determined
by geometrical features of the energy landscape.  We will return to
this important problem in Sec.\ \ref{energy_landscape}, but let us
first examine the discrete probabilities $\stateFrequency_k$ in
more detail.

\begin{figure}
\scalebox{0.4}{\includegraphics{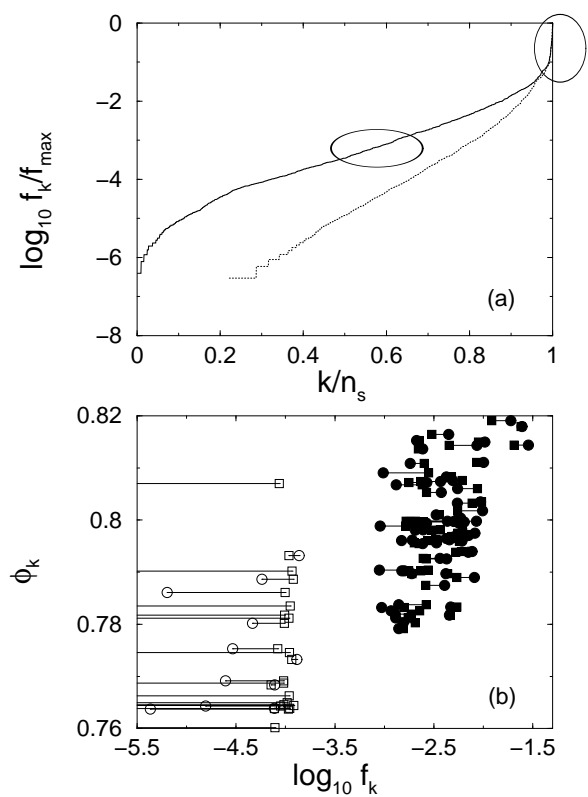}}%
\vspace{-0.18in}
\caption{\label{fig:frequent} (a) A comparison of the sorted discrete
probabilities $f_k$ normalized by $f_{\rm max}$ for the CG (solid
line) and MD (dotted line) methods including all MS packings for
$N=10$. $f_{\rm max}$ is the frequency of the most probable MS packing for
$N=10$.  (b) The discrete probabilities $f_k$ for the most frequent MS
packings depicted in (a) from the CG packing-generation method (filled
squares) are compared to the probabilities for the same packings
obtained from the MD method (filled circles).  A similar comparison of
the probabilities for less frequent packings in the packing-fraction
range $[0.76,0.82]$ obtained from the CG (open squares) and MD (open
circles) methods are also shown.  The circled regions in (a) identify
the MS packings that were compared in (b).}
\vspace{-0.22in}
\end{figure}


Figure \ref{sorted probabilities for different N} shows the discrete
probabilities $\stateFrequency_k$ for MS packings within a single
packing-fraction interval $\Delta\phi$ near the maximum of the density
of MS packings $\rho(\phi)$ for the CG method.  The probabilities
$\stateFrequency_k$ are sorted in increasing order and are plotted
versus the index $k$ normalized by the total number of states
$n_s(\Delta\phi)$ that exist in the interval.  The probabilities are
normalized by the maximal probability value $\stateFrequency_k^{\rm
max}$ (within $\Delta\phi$) and displayed on a logarithmic scale.
Results for systems with $N=10$, $12$, and $14$ particles are
provided.

\begin{figure}
\scalebox{0.4}{\includegraphics{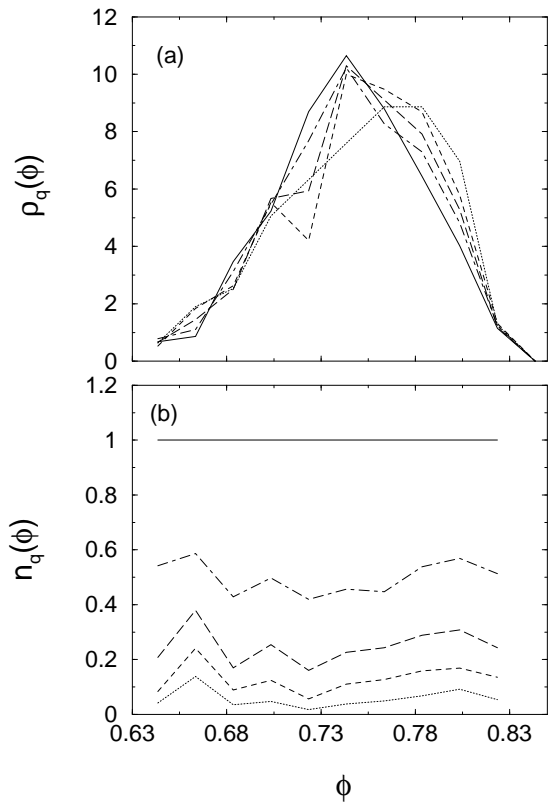}}
\vspace{-0.18in}
\caption{\label{distributions of most probable states} (a) The
truncated density of only the most frequent MS packings $\rho_q(\phi)$
for the CG energy minimization method in a $10$-particle system.  The
accumulated probability in each packing fraction interval $\Delta \phi
= 0.02$ is given by the parameter $q$; we display $q=0.3$ (dotted),
$0.5$ (dashed), $0.7$ (long-dashed), $0.9$ (dot-dashed), and $1.0$
(solid). (b) The fraction of distinct MS packings $n_q(\phi)$ that contribute 
to $\rho_q(\phi)$ in each $\phi$ interval for the same system
in (a).}
\vspace{-0.22in}
\end{figure}


It is important to notice that all three curves in Fig.\ \ref{sorted
probabilities for different N} have a similar shape.  Indeed, by
rescaling the data
\begin{equation}
\label{rescaled discrete probabilities}
\log_{10}(\stateFrequency_k/\stateFrequency_k^{\rm max})
  \to\alpha\log_{10}(\stateFrequency_k/\stateFrequency_k^{\rm max})
\end{equation}
by a factor $\alpha=O(1)$ the results for different system sizes
collapse onto a single master curve, as depicted in Fig.\ \ref{sorted
probabilities for different N} (b).  Since the total number of MS
packings in these three systems differs by more than two orders of
magnitude, the scaling \ref{rescaled discrete probabilities} is a
non-trivial result.  This result further reveals that there must be geometric
features of configuration space that determine the packing
probabilities.

We note that for $N=12$ and $14$ the total number of distinct packings
in the interval $\Delta\phi$ has not been measured directly---the most
infrequent states could not be generated because we were not able to
perform a sufficiently large number of trials with our current
computational resources.  The estimates for $n_s(\Delta\phi)$ used to
scale the horizontal axis in Fig.\ \ref{sorted probabilities for
different N} (and subsequent figures) were obtained by matching the
rescaled discrete probabilities for $N=12$ and $14$ to the
corresponding curve for $N=10$, for which the total number of MS
packings is known.

Unlike the density of MS packings, the packing probabilities
$\stateFrequency_k$ are protocol dependent.  To examine the effect of
different protocols on the probability distribution we compare, in
Fig.\ \ref{sorted probabilities for different protocols}({\it a\/}),
the sorted probabilities $\stateFrequency_k$ (scaled by
$\stateFrequency_k^{\rm max}$) for the CG and MD protocols in a
$10$-particle system over a small interval in packing fraction $\Delta
\phi$ near the peak in $\rho(\phi)$.  We observe that the relative
probabilities are significantly lower for the MD protocol, except in
the high-probability regime, where both sorted-probability curves
coincide.

The sensitivity of individual packing probabilities to the change of
the packing-generation protocol is examined in Fig.\ \ref{sorted
probabilities for different protocols}({\it b\/}).  In this figure we
compare the CG and MD probabilities of the $20$ packings that occur
most frequently in the interval $\Delta\phi$ according to the CG
protocol.  We also show $25$ packings that have much lower frequencies.
These results indicate that while the individual probabilities may
change even by several orders of magnitude (especially in the
low-probability regime), a significant shuffling between the sets of
frequent and infrequent packings does not occur when the energy
minimization protocol is changed.  The packings that are frequent
according to the CG protocol typically become even more frequent
according to the MD method, and the states that are infrequent become
even less frequent.  Moreover, the sets of the most frequent packings for
the two packing-generation methods nearly coincide.

Similar observations also hold for the set of {\it all} MS packings (rather
than only those within a small interval $\Delta\phi$), as can be seen
from the results shown in Fig.\ \ref{fig:frequent}.  In particular, we
find that for the $10$-particle system $\sim 80$ out of the $100$ most
frequent CG packings are also most frequent when they are generated
using the MD protocol.

We believe that the above results profoundly affect the way one should
interpret and explain a range of jamming and glassy phenomena
including random-close packing, dense slow granular flows, and arrested
dynamics in glass-forming liquids.  For example, the Edwards' entropy
description of nearly jammed granular materials is based on the
assumption that different jammed packings occur with approximately
equal weight.  Our findings indicate, however, that this assumption 
should be reconsidered.

Should we include all of the MS packings in statistical descriptions
of jammed and glassy systems even though the probabilities vary so
strongly or should we include only the most frequent packings?  If
only the most frequent packings are needed to describe jamming and
glassy phenomena, how do we distinguish between the frequent and
infrequent ones?  Similar questions apply to not only to Edwards'
entropy descriptions, but also to definitions of random close packing
and descriptions of glassy materials in terms of local minima in the
potential energy landscape.

Results from preliminary investigations aimed at addressing these
questions are shown in Fig.\ \ref{distributions of most probable
states}.  In panel (a), we plot the truncated density of MS packings
$\rho_q(\phi)$, which is defined as the density of only the most
frequent MS packings in each interval of $\phi$.  The accumulated
probability of these packings (normalized to unity in each
packing-fraction interval) is given by the parameter $q$.  For future
use, the set of all states contributing to $\rho_q$ is denoted by
$\probabilitySlab{q}$.  The results in Fig. \ref{distributions of most
probable states} are displayed for the CG packing-generation protocol
and $N=10$.

Consistent with the simulation data presented in Fig.\ \ref{sorted
probabilities for different N} for an interval near the maximum of the
density of MS packings, the most frequent 15\,\% of MS packings
contribute as much as 50\,\% of the local probability, according to
the results in Fig.\ \ref{distributions of most probable states} (b).
Also, to accumulate 90\,\% of the probability we only need roughly
50\,\% of the MS packings.

One of our key observations is that the density of MS packings is
insensitive to the cutoff probability $q$.  Thus, the peaks of
$\rho_q(\phi)$ for different probability-truncation levels nearly
coincide.  In contrast, the peaks of $\rho(\phi)$ and $P(\phi)$ in
Fig.\ \ref{fig:decomposition}({\rm a\/}) are more separated (although
still within the peak width), because the average frequency
distribution $\beta(\phi)$ depends exponentially on $\phi$.

The simulation results described so far reveal that the density of
discrete MS packings is an important quantity that may control, for
example, the position of the peak in the probability distribution of
MS packings for large systems, independent of the compaction
protocol.  However, we have also found that the probabilities of
individual MS packings vary dramatically from one packing to another.
We have also discovered interesting regularities of the packing
distributions, both as a function of $\phi$ and within narrow
packing-fraction intervals.

We believe that our observations are important for constructing
theoretical descriptions of amorphous packings and slow dynamics in
glassy materials.  Perhaps only the most frequent MS packings control
the behavior in these systems.  However, we must first understand what
determines the frequency with which MS packings occur in order to
identify correctly the set of frequent MS packings.  In the following
sections we investigate which structural and geometric factors play an
important role in determining the MS packing frequencies.

\section{potential energy landscape near MS packings}
\label{energy_landscape}

In this section we examine whether or not the probabilities of MS
packings can be correlated with local characteristics of the potential
energy landscape in the neighborhood of each packing.  The local
features include the heights of the energy barriers separating
different MS packings and the shape and size of the local region of
configuration space that is visited when mechanically stable packings
are subjected to thermal fluctuations.  As described below, we find
interesting correlations between the probabilities of the MS packings
and these local geometric features.  Our results, however, show that a
more global analysis of the topography of the potential energy
landscape is also required.

As in our packing-generation procedures, we consider a system
interacting via the one-sided soft spring potential \ref{spring
potential}.  However, our results can be applied more broadly.
First of all, local fluctuations around a MS packing in a system
interacting via a finite-range repulsive potential can approximately
be mapped onto the motion of a hard-disk system.  The effective disk
radii correspond to the distance at which the pair potential
$V(r_{ij})$ equals the thermal energy $T$ \cite{boltz}.  Thus, adding
thermal fluctuations to a MS packing is analogous to decompressing a
collectively jammed hard-disk system and then performing hard-particle
energy-conserving collision dynamics.

Second, crossing energy barriers and transitioning from one MS packing
to another resembles the evolution from the basin of one inherent
structure \cite{sastry} to another in glass-forming systems that
interact via continuous potentials.  We note that there are many more
inherent structures than MS packings at each $\phi$ so that a complete
enumeration becomes prohibitively costly even for extremely small
systems, unlike enumeration of the MS packings.  Also, MS packings can
be interpreted as metabasins \cite{vogel} of the PEL for systems with
finite-range repulsive potentials.  Thus our results are important not
only for understanding random-close packing but also glassy behavior
in soft- and hard-particle systems.

\subsection{Breaking times and energy barriers}

\subsubsection{Measurement of breaking times}

\begin{figure}
\scalebox{0.48}{\includegraphics{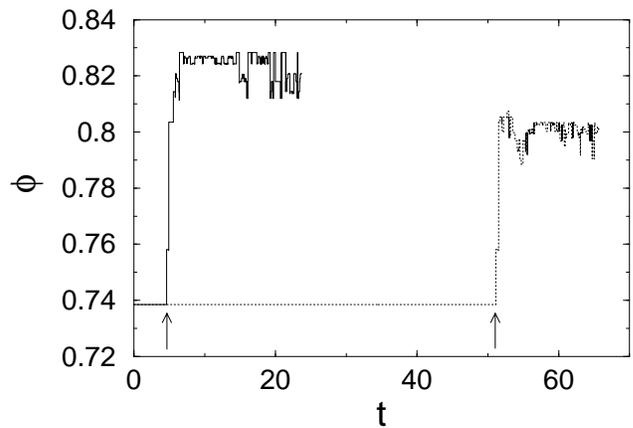}}
\vspace{-0.18in}
\caption{\label{brakup time measurements} 
The packing fraction $\phi$ of the MS packing that is nearest to the
instantaneous MD configuration at time $t$ (normalized by $t_{\rm
kin}$) for temperatures $T = 2.5 \times 10^{-6}$ (solid line) and a
factor of two lower (dotted line).  The minimal breaking times $\tau$ for these
two temperatures are indicated by the arrows.
}
\vspace{-0.22in}
\end{figure}


In our approach, we probe the local potential energy landscape near a
given MS packing via thermal fluctuations.  Each initial MS-packing
configuration is thermally excited by adding kinetic energy to the
system.  The energy is introduced by choosing the initial velocities
of the particles randomly from a Gaussian distribution with variance
$2 T$. We then allow the system to evolve at constant energy according
to the evolution equation \ref{newton1} with the damping coefficient
$b$ set to zero.  Even though the systems we study are small, they
behave like thermal systems in the sense that the $2T$ energy input is
quickly partitioned equally among the configurational and kinetic
degrees of freedom: $\langle V \rangle \sim T$ and $\langle K \rangle
\sim T$, where $K$ is the kinetic energy per particle.

During the course of the molecular-dynamics run, we periodically save
the particle coordinates.  For each snapshot, the particle coordinates
are fed into our MS packing-generation routine to find the nearest MS
packing using the CG energy minimization scheme.  The nearest
mechanically stable packings for each snapshot are then compared to
the original MS packing at $t=0$.

This procedure allows us to measure the first-passage breakup time
$\tau$ for the system to make a transition from the original MS
packing to a different one.  In this section, all times are measured
in units of the kinetic timescale $t_{\rm kin} = \sigma
\sqrt{m/T}$. We chose the time interval between instantaneous MD
snapshots to be $5$--$20~\Delta t \ll \tau$ (where $\Delta t$ is the
integration time step), which is small enough so that it does not
influence our results.  For each initial MS packing and temperature
$T$, the procedure is repeated at least $20$ times with random initial
velocities to obtain the average breaking time $\langle \tau \rangle$.

A typical example of the breaking-time measurements at two different
temperatures for a MS packing near the peak of the density of states
$\rho(\phi)$ is depicted in Fig.\ \ref{brakup time measurements}.  The
figure shows the evolution of the packing fraction $\phi$ of the MS
packing that is nearest to each instantaneous MD configuration.  After
the state breaks away from the neighborhood of the initial MS packing,
it cascades through a set of MS packings with gradually increasing
$\phi$ and ends up oscillating between several high-packing-fraction,
high-probability MS packings.  We emphasize that the molecular
dynamics evolution takes place at {\it constant $\phi$}.  The separate
compression/decompression and energy-minimization procedure is used
only to identify the corresponding MS packing for each instantaneous
configuration from the MD evolution.

Our method for finding the nearest MS packings is similar to thermally
quenching instantaneous MD configurations to the nearest inherent
structure or local potential energy minimum \cite{sastry}.  However,
apart from energy minimization, in our procedure we incorporate the
additional steps of shrinking and growing the particles to arrive at a
MS packing with infinitesimal particle overlaps \cite{sw}.

\begin{figure}
\scalebox{0.5}{\includegraphics{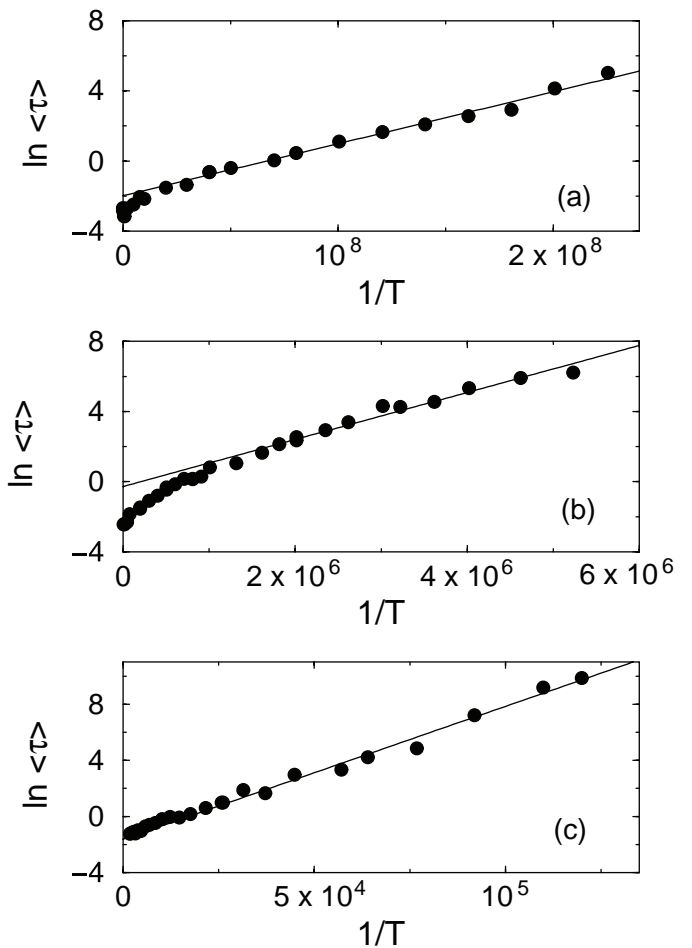}}%
\vspace{-0.18in}
\caption{\label{fig:arrhenius} The average breaking time $\langle \tau \rangle$
required for the system to surmount the lowest energy barrier
$\epsilon_0$ versus inverse temperature $1/T$ for three different MS
packings with $N=10$ particles.  The slope of the solid lines in
panels (a)-(c) are $\epsilon_0 \approx 3.3 \times 10^{-8}$, $1.4
\times 10^{-6}$, and $1.1 \times 10^{-4}$.}
\vspace{-0.22in}
\end{figure}


\subsubsection{Measurement of energy barrier heights}

In order to determine the height of the energy barrier that separates
a given MS packing from other packings, we have performed a series of
breaking-time measurements over a wide range of temperatures $T$.  The
simulations were performed on a randomly selected group of $105$ MS packings
with $N=10$.

Measurements of the average minimal breaking time $\langle\tau\rangle$ versus
the inverse temperature for three different MS packings are presented
in Fig.\ \ref{fig:arrhenius}. Each data point in these plots was
obtained from an average of at least 20 independent measurements of
the breaking time; the standard deviation of $\tau$ is comparable to
the symbol size.

The results in Fig.\ \ref{fig:arrhenius} indicate that for
sufficiently low temperatures the system exhibits Arrhenius behavior
\begin{equation}
\label{arrhen}
\langle \tau \rangle = \tau_{\infty} e^{\epsilon_0/T}.
\end{equation}
The quantity $\epsilon_0$ in the above relation is interpreted as the
height of the energy barrier that separates a given MS packing from
others.  Below, both $\epsilon_0$ and $T$ will be measured in units of 
the characteristic energy scale $\epsilon$ of the repulsive spring potential
in Eq.~\ref{spring potential}. 
The parameter $\tau_\infty$ characterizes the
timescale at which the system explores the energy landscape.

We find from our results that the exponential behavior \ref{arrhen}
is quite robust despite the fact that we use small systems with only
$\sim 20$ translational degrees of freedom.  The evolution occurs at
constant energy with no heat reservoir---yet we observe
pronounced thermal behavior.  We note that the breaking time in the
Arrhenius regime varies by several orders of magnitude, which allows us
to determine the barrier height $\epsilon_0$ quite accurately.

At high temperatures, the MS packings break to a number of different
destination MS packings because the system possesses enough kinetic
energy to traverse a broad range of barriers.  In the low-temperature
regime, the system breaks by jumping over the lowest energy barrier,
unless there are several barriers of nearly the same magnitude.  By
fitting the simulation data (such as those presented in
Fig.~\ref{fig:arrhenius}) to the Arrhenius form \ref{arrhen} in the
low-temperature regime we can thus measure the lowest energy barrier
$\epsilon_0$ for each MS packing studied.  At moderate temperatures,
we can separately calculate the average breaking time $\langle \tau
\rangle$ for each destination MS packing, which, in principle, allows
us to measure the lowest, next lowest, and subsequent higher energy
barriers.

\begin{figure}
\scalebox{0.4}{\includegraphics{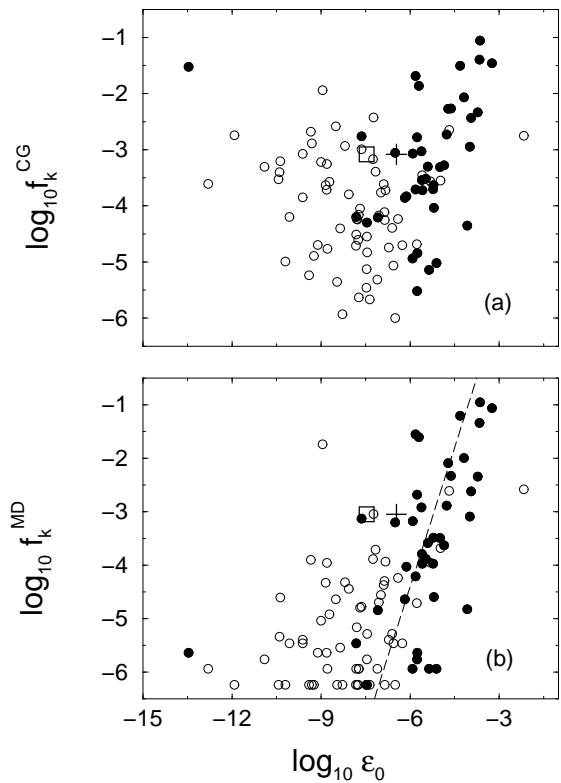}}
\vspace{-0.18in}
\caption{\label{barrier hight CG and MD} 
The discrete probability $f_k$ with which a given MS packing occurs
versus the lowest energy barrier $\epsilon_0$ associated with that
packing using the (a) CG and (b) MD energy minimization methods.  All
$105$ MS packings in our sample are included.  The filled circles
correspond to the most probable set of MS packings
$\probabilitySlab{q=0.7}$, while the open circles indicate the
remaining less frequent ones.  The long-dashed line in panel (b) with
slope $1.7$ points out the correlation between $\stateFrequency_k^{\rm
MD}$ and $\epsilon_0$.  The open square (cross) in both panels
corresponds to the twin packing (meta-packing) discussed in
Sec.~\ref{relation}.
}
\vspace{-0.22in}
\end{figure}


\subsubsection{Relation between energy barriers an probabilities of MS
packings}
\label{relation}

The Arrhenius plots presented in Fig.\ \ref{fig:arrhenius} reveal that
the magnitudes of the minimal energy barriers $\epsilon_0$ separating
distinct MS packings vary by many orders of magnitude. It is thus
interesting to determine whether or not there is a relation between the
energy-barrier heights and the MS packing probabilities
$\stateFrequency_k$, which exhibit a similar large variation, as discussed
in Sec.\ \ref{Packing frequency distribution}.  We have measured the
energy-barrier heights for a sample of $105$ MS packings, which comprise
only about $6.5\%$ of the total number of MS packings for the
$10$-particle system considered (cf.\ the results in Table
\ref{tab:table1}).  The small sample size does not allow for a
complete quantitative analysis of the problem, but we are able to make
some interesting qualitative observations.

In Fig.~\ref{barrier hight CG and MD} we compare the minimal energy
barrier heights $\epsilon_0$ and the corresponding probabilities
$\stateFrequency_k$ for our sample of MS packings.  The probabilities
shown in panel ({\it a}\/) were obtained using the CG protocol, and
those depicted in panel ({\it b}\/) correspond to the MD algorithm.
In both plots, the filled symbols indicate the data points for the
packings in the set $\probabilitySlab{q}$ of the most probable
packings contributing to the truncated density of states
$\rho_q(\phi)$ at the probability truncation level $q=0.70$.  (The
set $\probabilitySlab{q}$ is defined near the end of Sec.\
\ref{Numerical results for frequency distributions}.)  The open 
symbols correspond to less probable packings.

There are several interesting features to be noticed in these plots.
In the case of the CG protocol, illustrated in Fig.\ \ref{barrier
hight CG and MD}({\it a}\/), we see only a gross correlation between
$\stateFrequency_k$ and $\epsilon_0$.  The high-probability states
tend to have high energy barriers, and the barriers of low-probability
packings do not exceed $10^{-5}$ in our sample.
Otherwise, the scatter in the data is very large; for example, the
energy barriers of the packings in the probability range
$10^{-4}$ to $10^{-3}$ vary by nearly ten orders of magnitude.

In contrast, there is a much stronger correlation between the
probabilities and energy barriers in the subset of packings that have
the maximal probability in each local packing-fraction region.  We
note that for a given value of $\stateFrequency_k$, the packings from
the high-probability subset $\probabilitySlab{q=0.7}$ (shown
as filled circles in Fig.\ \ref{barrier hight CG and MD}({\it a}\/) )
tend to have the highest energy barriers $\epsilon_0$ for
a given $\stateFrequency_k$.

The results presented in Fig.\ \ref{barrier hight CG and MD}({\it
b}\/) show that the points in the high-probability subset are
insensitive to the packing-generation protocol, consistent with the
results shown in Fig.\ \ref{distributions of most probable states}.
However, the packings outside the subset $\probabilitySlab{q=0.7}$ shift
significantly in the low-probability direction when switching from the
CG to MD method.  Since packings outside $\probabilitySlab{q=0.7}$ tend to
have low energy barriers, the MD probabilities are much more strongly
correlated with the heights of the minimal energy barriers than the CG
probabilities.  In fact, the data for the most probable states
$\probabilitySlab{q=0.7}$ in Fig.\ \ref{barrier hight CG and MD}({\it
b}\/) roughly scale as a power-law $\stateFrequency_k^{\rm
MD} \sim \epsilon_0^{\lambda}$ with $\lambda \approx 2$.

The above observations point out that the relationship between the
probabilities of MS packings and $\epsilon_0$ is complex.  On the one
hand, the magnitude of the minimal energy barrier separating a given
MS packing from other packings is not directly linked to the packing
probability $\stateFrequency_k$, especially for the CG protocol.  On
the other hand, the energy-barrier heights clearly determine
some important features of the probability distribution.  In
particular, the probabilities $\stateFrequency_k$ of MS packings with
low energy barriers depend strongly on the packing-generation
protocol, which can be seen by comparing the results in Figs. \ref{barrier
hight CG and MD} ({\it a\/}) and ({\it b\/}).  Moreover, there is a
significant correlation between $\epsilon_0$ and $\stateFrequency_k$
for a subset of the most frequent packings for each $\phi$.

Strong protocol dependence of the probabilities $\stateFrequency_k$
for MS packings with low values of $\epsilon_0$ can be understood
intuitively by noting that a system with weak but nonvanishing
thermal fluctuations is able to surmount low energy barriers but not
high ones.  Thus for protocols such as the MD algorithm, the
low-barrier packings tend to have low probabilities.  After traversing
the low barriers, the system becomes trapped in packings with higher
barriers, which thus have higher probabilities.

We expect similar behavior to occur during the relaxation process of
glass-forming liquids following a thermal quench.  During the quench,
the system first quickly relaxes by surmounting low energy barriers.
At later times it evolves much more slowly through a set of
high-barrier states, which is closely related to the high-probability
subset $\probabilitySlab{q}$ of MS packings.  During the
slow-evolution stage, the properties of the system are thus strongly
affected by the truncated density of the most frequent states similar
to the truncated density of MS packings $\rho_q(\phi)$ introduced in
Fig.\ \ref{distributions of most probable states}.

The large scatter of the data points in Fig.\ \ref{barrier hight CG
and MD} indicates that additional geometrical features of the
potential energy landscape---not simply the minimal energy
barriers---must influence the probabilities of MS packings.  Below we
qualitatively discuss a few nonlocal topographic features that may
play a significant role.  A more detailed, quantitative investigation
of such features is beyond the scope of the current study, but will 
be pursued in future work.

\begin{figure}
\scalebox{0.4}{\includegraphics{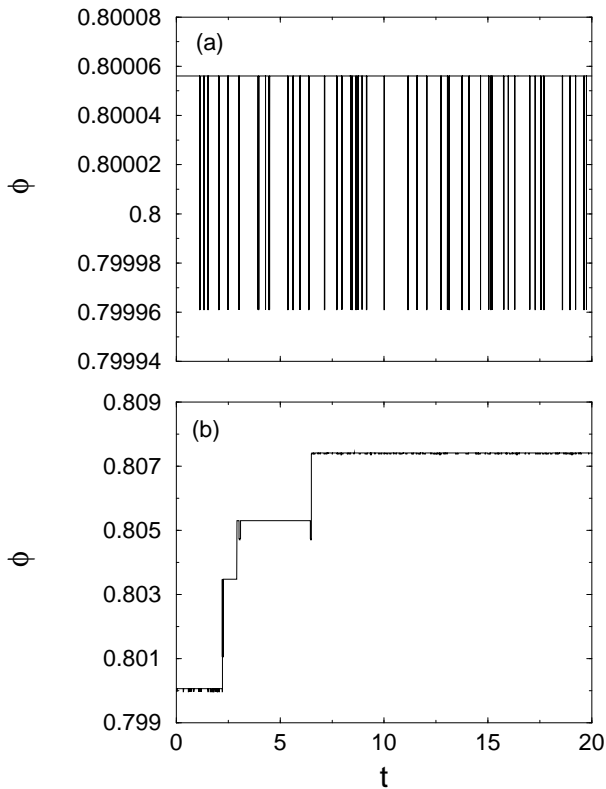}}%
\vspace{-0.18in}
\caption{\label{twin state} The packing fraction $\phi$ of the MS packing
that is nearest to the instantaneous MD configuration at time $t$ for
(a) a temperature comparable to the minimal energy-barrier height 
$T \sim \epsilon_0$ and (b) a temperature 
that is four times larger.}
\vspace{-0.22in}
\end{figure}


\paragraph*{Twin packings and meta-packings}

A relatively simple feature of the PEL that can give rise to
significant fluctuations in the relation between the minimal energy
barrier $\epsilon_0$ and the probabilities $\stateFrequency_k$ is
illustrated in Fig.\ \ref{twin state}.  This figure shows two sample
breakup trajectories for the MS packing indicated by the open square
in Fig.\ \ref{barrier hight CG and MD}.  The trajectory depicted in
Fig.\ \ref{twin state}({\it a}\/) corresponds to an initial thermal
excitation with $T$ comparable to the minimal barrier height
$\epsilon_0$; the trajectory depicted in Fig.\ \ref{twin state}({\it
b}\/) corresponds to a thermal excitation that is four times larger.

According to the results in Fig.\ \ref{twin state}, at the lower
temperature the system oscillates between two MS packings that have
nearly the same packing fraction $\phi$.  We refer to these as {\it
twin packings}.  In order for the system to make a transition to a
non-twin packing within this timescale, the temperature must be
increased dramatically.  After surmounting the barrier that takes the
system to a non-twin packing, the system undergoes the usual cascade
of transitions through a sequence of packings with increasing $\phi$
(cf.\ the trajectory depicted in Fig.\ \ref{brakup time
measurements}).

The meta-packing that consists of the two twin packings between which
the system oscillates at low temperatures can have a much higher
energy barrier than the barrier $\epsilon_0$ separating the two
component packings.  We expect that the probability of the twin
packings is controlled by the meta-packing energy barrier $\epsilon_1$
rather than the minimal barrier $\epsilon_0\ll\epsilon_1$.  Thus, we
have also included the data point $(\epsilon_1,\stateFrequency_k)$ for the twin-packing
in Fig.\ \ref{barrier hight CG and MD} using a cross as the symbol.
Note that using $\epsilon_1$ instead of $\epsilon_0$ brings this point
closer to the cluster of data for the most probable subset of packings
$\probabilitySlab{q=0.7}$.  A generalization of this picture to a
larger group of packings will be carried out in a future publication
\cite{future}.  When we perform this analysis we may likely find that
more than two component packings can belong to a given meta-packing.
We note that the concept of meta-packings is closely related to that
of metabasins of the PEL \cite{debenedetti,vogel}.

\paragraph*{Probability streams}

Identifying neighboring packings that are separated by small energy
barriers and grouping them into meta-packings may decrease the scatter
of the data in Fig.\ \ref{barrier hight CG and MD}, but it is unlikely
to eliminate it.  Thus, it is apparent that not
only energy barriers but other geometric features of the PEL are
important for determining the probabilities $\stateFrequency_k$.  At
present, we do not have any direct measurements of these additional
features.  However, our simulation results provide important clues
that indicate directions for further investigations of the problem.

We may consider here two possibilities.  First, one could assume that
the volume $\Omega_k$ of the local region in configuration space
with energy $V<\epsilon_0$ in the neighborhood of a given MS packing
determines its probability $\stateFrequency_k$.  For a given
$\epsilon_0$, this volume may have large variations due to variations
of the shape of the local potential-energy basin from one MS packing
to another.  However, as shown in Sec.\ \ref{Section on displacement
matrix} below, this possibility can be excluded since the
multi-dimensional volume $\Omega_k$ depends too strongly on the
height of the energy barrier.

Thus, a more complex, non-local scenario may better explain the MS
packing probabilities.  According to this scenario, the probability
$\stateFrequency_k$ of a MS packing depends on two competing factors:
the probability of getting into the neighborhood of that packing and
the probability of leaving the local region before the MS packing is
reached during the relaxation process.  The probability of leaving the
local region is largely controlled by the height of the energy
barrier, and thus it is strongly protocol dependent.  The probability
of arriving into the neighborhood of an MS packing is affected by
the size of the local region (therefore states from the high
probability subset $\probabilitySlab{q}$ tend to have high energy
barriers), but it is more strongly influenced by the chain of events
that brought the system into the neighborhood of the MS packing.
These events, in turn, are determined by the features of PEL that can
be far away from the local potential-energy basin for a given MS
packing.

The large scatter in the probabilities for a given value of the energy
barrier indicates that the flux of probability during the compaction
process is very nonuniform.  It appears that there are many ``dry''
regions, with very small probability flux, and ``probability
streams,'' where the probability flux is very large.  If a given MS
packing is in the path of such probability streams, it may have a very
large probability, even if the energy barrier $\epsilon_0$ associated
with this packing is low.

We note that in our intuitive picture, probability flux is akin to
water flow in a rugged mountainous landscape.  This analogy, however,
may be oversimplified due to the highly multidimensional
character of the PEL for particulate systems.

\begin{figure}
\scalebox{0.5}{\includegraphics{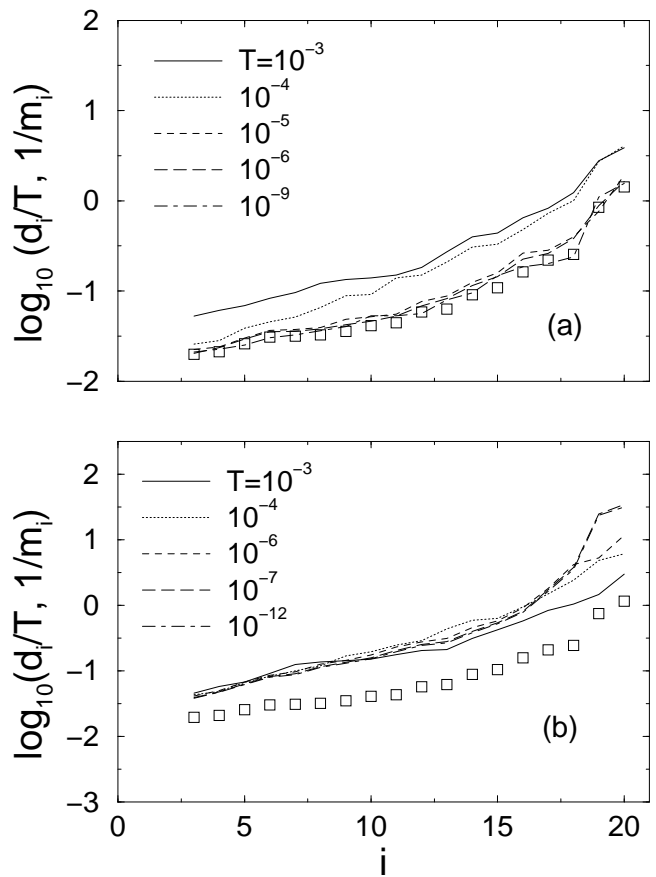}}
\vspace{-0.18in}
\caption{\label{one-sided springs} 
Comparison of the eigenvalues of the displacement matrix $d_i/T$ for
several temperatures $T$ (curves) and those of the inverse dynamical
matrix $1/m_i$ (open squares) for (a) a system that has been
compressed by $1\%$ from the jamming threshold and (b) a MS
packing.  Note that when plotting the eigenvalues of the dynamical and
displacement matrices, we include only the $dN - d$ nontrivial
eigenvalues and order them from smallest to largest.}
\vspace{-0.22in}
\end{figure}


\subsection{Displacement fluctuations}
\label{Section on displacement matrix}

\subsubsection{Displacement and dynamical matrices}

After a digression on possible non-local features of the PEL that can
control the large fluctuations in the probabilities of MS packings, we
now return to our analysis of local features of the PEL near a given
MS packing.  In this section we focus on measuring the shape and size of the
local region visited by the system when the MS packing is thermally
excited.

In general, the local shape of the PEL basin can be studied using both
static and dynamic techniques.  The static method relies on an
analysis of the eigenvalue spectrum of the dynamical (Hessian) matrix
\ref{offdiagonal} and \ref{diagonal}.  The dynamic technique that is
applied in this section is based on the evaluation of the
$2N$-dimensional matrix of displacements away from the reference MS
packing while the system fluctuates after input of thermal energy.
The displacement matrix is defined as
\begin{equation}
\label{displacement}
D_{i\alpha,j\beta} = \langle \left( r_{i\alpha} - r_{i\alpha 0} \right)
\left( r_{j\beta} - r_{j\beta 0}\right) \rangle,
\end{equation}
where $r_{i \alpha 0}$ are the coordinates of the reference MS
packing, and $r_{i\alpha}$ are the current coordinates.  Again, the
indices $i$ and $j$ refer to the particles, and $\alpha,\beta=x,y$
represent the Cartesian coordinates. The average in Eq.\
\ref{displacement} is taken both over times $t<\tau$ less than the
minimal breaking time and over at least $20$ realizations at each $T$,
weighted by the corresponding breaking time $\tau$.

In harmonic systems, the eigenvalues of the displacement matrix $d_i$ are 
trivially related to those of the dynamical matrix $m_i$
\begin{equation}
\label{harmonic}
d_i = \frac{T}{m_i}.
\end{equation}
Near an MS packing, however, our system of disks is strongly
anharmonic, due to the one-sided character of the spring potential
\ref{spring potential} at the jamming threshold.  Thus, relation
\ref{harmonic} is not guaranteed to hold even in the low-temperature
limit.

\begin{figure}
\scalebox{0.45}{\includegraphics{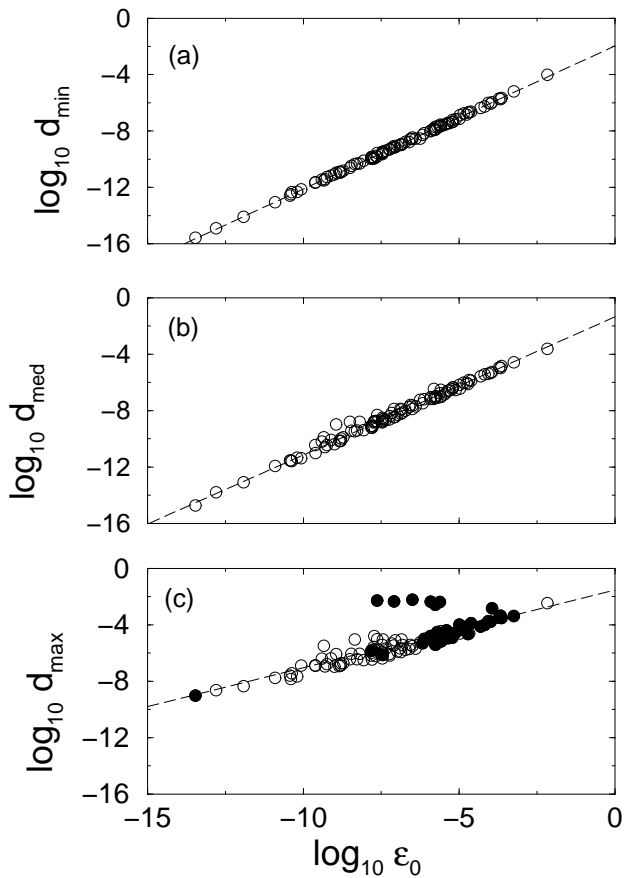}}
\caption{\label{fig:eigens}
(a) The minimum $d_{\rm min}$, (b) median $d_{\rm med}$, and (c)
maximum $d_{\rm max}$ eigenvalues of the displacement matrix plotted
versus the minimal energy-barrier height $\epsilon_0$ at temperature
$T/\epsilon_0 = 0.2$ for $N=10$ and all $105$ packings in our sample.
The eigenvalues of the displacement matrix scale as $d_i \sim
\epsilon_0^{\gamma_i}$, and the power-law exponents $\gamma_i$ are
given in Fig.~\ref{fig:scaling_exponent}. For $d_{\rm max}$, a few
packings in the subset $\probabilitySlab{q=0.7}$ (solid circles) deviate
from the main trend. }
\end{figure}

We find that for compressed systems with finite particle overlaps
there is a characteristic temperature $T_c$ below which relation
\ref{harmonic} is satisfied.  For example, a system compressed by
$1\,\%$ above the jamming threshold behaves harmonically for $T < T_c
\approx 10^{-5}$, as illustrated in Fig.~\ref{one-sided springs}({\it
a\/}).  However, the temperature $T_c$ tends to zero as the amount of
overlap decreases; therefore relation \ref{harmonic} is not valid
for MS packings even at infinitesimal temperatures.  The quantities on
the left and right hand sides of Eq.\ \ref{harmonic} can differ by
more than an order of magnitude for MS packings as shown in Fig.\
\ref{one-sided springs}({\it b\/}).  

In what follows, we focus on the displacement matrix
\ref{displacement} (rather than the dynamical matrix) because it
more reliably measures the local shape of potential energy landscape
near MS packings.  Moreover, at finite temperatures the matrix
$D_{i\alpha,j\beta}$ yields information about the entire region of
configuration space with energy $V<\epsilon_0$ near a given MS
packing.  Note that since we are considering thermally fluctuating
systems in which all particles move, the distinction between
rattler and non-rattler particles becomes less important and we will
not distinguish between these two types of particles in the following
discussion.

\subsubsection{Eigenvalues of displacement matrix}
\label{Section on eigenvalues of displacement matrix}

The results displayed in Fig.\ \ref{one-sided springs} (b) show that
the eigenvalue spectrum of the displacement matrix for MS packings can
be highly nonuniform.  In this example, the ratio of the largest to
smallest displacement eigenvalues $d_{\rm max}/d_{\rm min}$ is
approximately $10^3$ at low temperatures.  It is thus interesting to
examine how  this ratio, and more generally the shape of the basin
near each MS packing, varies from one packing to another.

The shape of the local basins was explored by evaluating the
displacement matrix \ref{displacement} for different MS packings at
a fixed value of $T/\epsilon_0$, so that for each packing we study
comparable relative displacements away from potential energy minimum.
We find that the eigenvalues of the displacement matrix exhibit 
power-law scaling with the energy-barrier height,
\begin{equation}
\label{gamma}
d_i \sim \epsilon_0^{\gamma_i},
\end{equation}
where the scaling exponent $\gamma_i$ depends on the position $i$ in
the set of $d(N-1)$ eigenvalues ordered by their magnitude.  

\begin{figure}
\scalebox{0.43}{\includegraphics{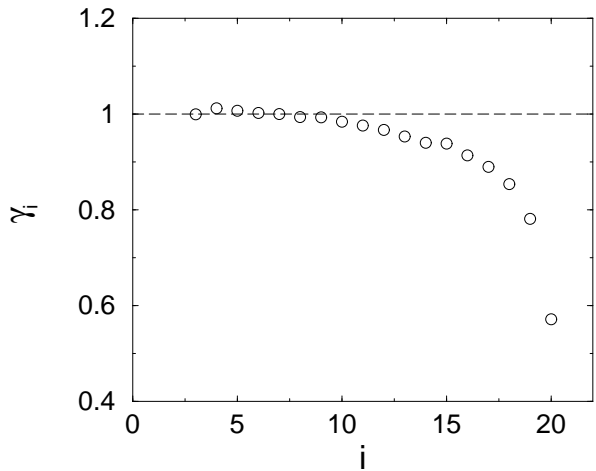}}
\caption{\label{fig:scaling_exponent}
The power-law exponents $\gamma_i$ that determine how the eigenvalues of the
displacement matrix scale with the minimal energy-barrier height
$\epsilon_0$. The index $i$ runs from the smallest eigenvalue of the
displacement matrix to the largest. For harmonic systems, all
$\gamma_i = 1$.}
\end{figure}

The power-law behavior \ref{gamma} is illustrated in
Fig.~\ref{fig:eigens} for the minimum, median, and maximum eigenvalues
of the displacement matrix for $N=10$ at fixed $T/\epsilon_0 = 0.2$.
We also performed measurements at $T/\epsilon_0 = 0.3$ and obtained
similar results.  The power-laws extend over at least $10$ orders of
magnitude in $\epsilon_0$, although there is some scatter in the data
for the largest eigenvalue $d_{\rm max}$.  $d_{\rm max}$ for several
of the packings from the most frequent subset
$\probabilitySlab{q=0.7}$ deviate from the main trend.

The scaling exponents are given in Fig.~\ref{fig:scaling_exponent} and
range from approximately $0.6$ for the largest eigenvalue to near $1.0$
for eigenvalues below the median.  The fact that the largest several
scaling exponents differ significantly from unity again indicates that
MS packings are anharmonic, since in harmonic systems $\gamma_i = 1$
(assuming that the eigenvalues of the dynamical matrix are independent
of $\epsilon_0$).

\begin{figure}
\scalebox{0.5}{\includegraphics{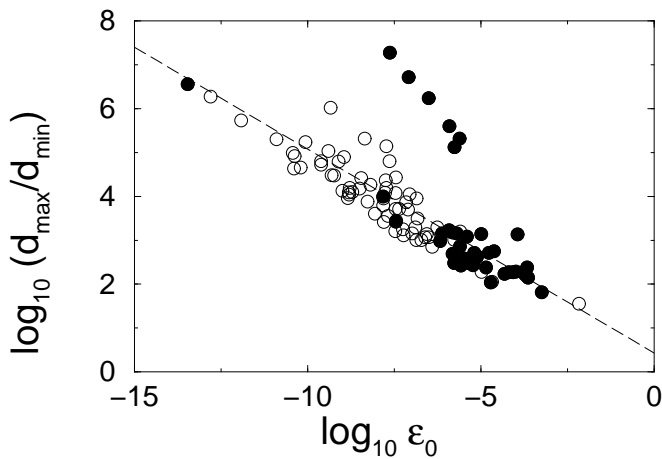}}
\caption{\label{fig:ratio_e}
Ratio $d_{\rm max}/d_{\rm min}$ of the maximum and minimum eigenvalues
of displacement matrix versus the minimal energy-barrier height
$\epsilon_0$ at temperature $T/\epsilon_0 = 0.2$ for $N=10$ and all
$105$ packings in our sample. The long-dashed line has slope 
$-0.5$. The most frequent packings in the subset $\probabilitySlab{q=0.7}$
are indicated by filled circles while the rest are indicated by 
open circles.}
\end{figure}

To illustrate the non-uniformity of the displacement-matrix spectrum,
in Fig.~\ref{fig:ratio_e} we plot the ratio $d_{\rm max}/d_{\rm min}$
versus $\epsilon_0$ at fixed $T/\epsilon_0 = 0.2$.
We find that the ratio roughly obeys the power-law
\begin{equation}
\label{rat}
\frac{d_{\rm max}}{d_{\rm min}} \sim \epsilon_0^{-0.5},
\end{equation}
consistent with the results shown in Figs.\ \ref{fig:eigens} (a) and
(c) and \ref{fig:scaling_exponent}.  Fig.~\ref{fig:ratio_e} emphasizes
that MS packings can possess extremely nonuniform displacement
fluctuations with $d_{\rm max}/d_{\rm min}$ as large as $10^7$
for packings with small energy barriers.  As in Fig.\ \ref{fig:eigens}
(c), there are several frequent MS packings with $\epsilon_0 <
10^{-5}$ that have considerably larger ratios than this trend, but the
vast majority of points obey (\ref{rat}).

Our data shows that MS packings possess highly nonuniform displacement
eigenvalue spectra (nearly all have $d_{\rm max}/d_{\rm min} > 100$)
and the non-uniformity substantially increases with decreasing
$\epsilon_0$.  This suggests that the basins associated with MS
packings are highly anisotropic in configuration space, and this gives
rise to displacement fluctuations that are much larger in one or a few
directions than others.  Thus, one simple picture is that MS packings
break only along particular directions in configuration space and that
the breaking direction will be correlated with the direction in which the
displacement fluctuations are the largest. We will investigate this
intuitive picture in future studies.

The results presented in Figs.\ \ref{fig:eigens}--\ref{fig:ratio_e}
allow us to determine if there is a direct link between the volume
$\Omega_k$ and the packing probabilities $\stateFrequency_k$,
where $\Omega_k$ is the volume of configuration space near a given
MS packing that contains points with potential energy $V<\epsilon_0$ .
An estimate of this volume can be obtained by assuming that at
temperature $T\sim\epsilon_0$ the system explores a large portion of
$\Omega_k$.  Accordingly, we find
\begin{equation}
\label{size of local region}
\Omega_k\sim\prod_{i=1}^{d(N-1)}\sqrt{d_i} 
  \sim \epsilon_0^\eta,
\end{equation}
where $\eta=\frac{1}{2}\sum_{i=1}^{d(N-1)} \gamma_i$.  The last
expression in (\ref{size of local region}) was obtained using the
relation between $d_i$ and $\epsilon_0$ in (\ref{gamma}).

Thus, if we assume that the packing probabilities are
controlled by the volume $\Omega_k$, we obtain 
\begin{equation}
\label{final}
\stateFrequency_k \sim \epsilon_0^\eta,
\end{equation}
where $\eta\approx 7$ for the $10$-particle system considered
here.  However, our results in Fig.~\ref{barrier hight CG and MD} show
that the dependence of the packing probabilities on $\epsilon_0$ is
much weaker than that predicted by (\ref{final}).  The exponent for
the $10$-particle data is approximately $2$, not $7$.

Thus, our results indicate that the packing probabilities are
determined by a much lower dimensional quantity than the volume
$\Omega_k$.  Our results are consistent with a scenario in which
the packing frequencies are determined by only the largest several
displacement eigenvalues.  It is likely that the packing probabilities
are correlated with the largest displacement eigenvalues because they 
may correspond to the (initial) escape directions from the basin
of a MS packing. 

\begin{figure}
\scalebox{0.6}{\includegraphics{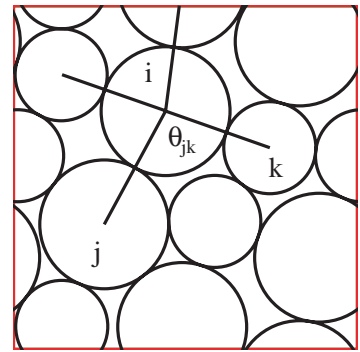}}%
\vspace{-0.05in}
\caption{\label{fig:freq} Definition of contact angles.  $\theta_{jk}$
is the angle between the lines that connect the central particle $i$
to a pair of adjacent, contacting neighbors $j$ and $k$. There are
four contact angles associated with particle $i$.}
\vspace{-0.22in}
\end{figure}


\section{Structural properties of MS packings}
\label{structure}

We also examined structural properties of MS packings in an attempt 
to identify features that control MS packing probabilities.
Specifically, we measured the probability distributions for the angles
between lines connecting centers of particles in contact (cf., the
definition in Fig.~\ref{fig:freq}).  We focused on the distribution of
maximal angles between particle contacts because large angles (close
to $180^\circ$) correspond to unsupported (and thus unstable) chains
of nearly collinear particles.

\begin{figure}
\scalebox{0.4}{\includegraphics{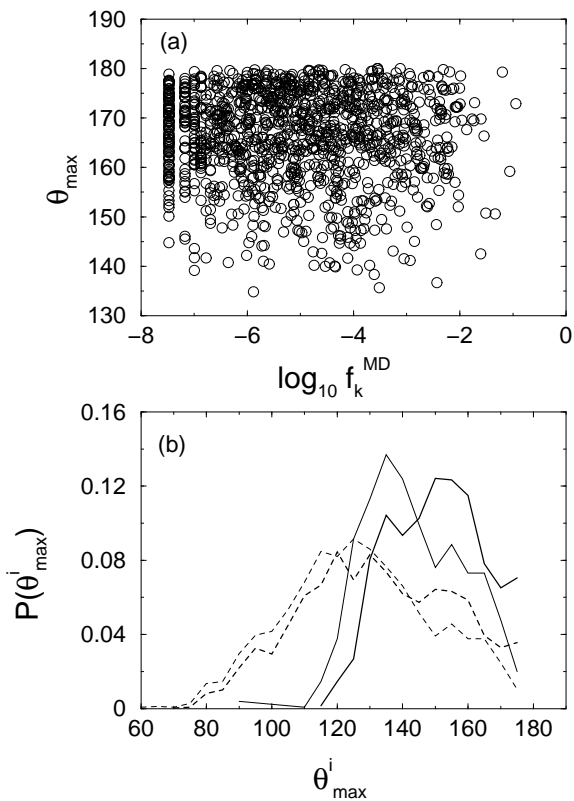}}%
\vspace{-0.05in}
\caption{\label{fig:angle} (a) Maximum contact angle $\theta_{\rm
max}$ (in degrees) for each MS packing versus the frequency $f_k^{\rm
MD}$ with which it occurs for a $10$-particle system using the MD
energy minimization method. (b) The distribution of the largest
contact angle $\theta^i_{\rm max}$ (in degrees) on each particle for
the $100$ most frequent (thin lines) and $100$ most infrequent (thick
lines) for the same system in (a).  The dashed (solid) lines include
all (larger half) $\theta^i_{\rm max}$ in each configuration.}
\vspace{-0.22in}
\end{figure}


As defined in Fig.~\ref{fig:freq}, $\theta_{jk}$ is the angle between
the lines connecting the central particle $i$ to pairs of adjacent,
contacting neighbors $j$ and $k$.  A given particle will possess $n_c$
contact angles, where $n_c$ is the number of contacts for that
particle.  We calculated the maximum contact angle $\theta^i_{\rm
max}$ for each of the $N'$ non-rattler particles and the maximum angle
$\theta_{\rm max}$ for each configuration.

In Fig.~\ref{fig:angle} (a), we present a scatter plot of $\theta_{\rm
max}$ for each MS packing in a system with $N=10$ particles versus the
frequency $\stateFrequency_k$ with which the packing occurs for the MD
packing-generation protocol.  This plot shows that there is no strong
correlation between $\theta_{\rm max}$ and $\stateFrequency_k$.  In
fact, some highly probable MS packings possess large contact angles
near $180^\circ$.  

Unsupported nearly linear chains of particles imply a low energy
barrier.  Since the barrier heights are correlated with the
probability $\stateFrequency_k$ for the MD protocol, our results may
thus suggest an important role of near contacts between particles.
Such contacts do not directly support the chain but they may prevent
transitions to other MS packings when the system is fluctuating
\cite{donev_com}. We will return to this problem in future investigations.

We also measured the distribution of the largest contact angles on
each particle $\theta^i_{\rm max}$.  In Fig.~\ref{fig:angle} (b), we
compare the distributions of $\theta^i_{\rm max}$ for the $100$ most
frequent (thin lines) and $100$ most infrequent (thick lines) MS
packings generated for a $10$-particle system using the MD protocol.
The figure shows two sets of curves.  The probability distribution
represented by the solid lines includes all $\theta^i_{\rm max}$ in the
system; the dashed lines represent the distribution of only the
largest half of the angles for each MS packing.

In contrast to the results represented in Fig.~\ref{fig:angle} (a),
the distributions depicted in Fig.\ \ref{fig:angle} (b) do show a
clear correlation between large angles $\theta^i_{\rm max}$ and the
frequencies $\stateFrequency_k$ of MS packings: The infrequent ones
have an excess of large angles near $160^{\circ}$ compared to the
frequent packings.  Thus infrequent MS packings tend to have multiple
particles with large contact angles.

The structural differences between the frequent and infrequent states
that we see in Fig.\ \ref{fig:angle} are quite subtle.  We do not have
any significant correlation with the single angle $\theta_{\rm max}$
in the packing.  The differences show up, however, collectively in the
distribution of the contact angles for each particle in the packing.
One of our long-term goals is to connect important geometrical
features of configuration space to structural properties of MS
packings that can be measured in experiments.

\section{Is there a hidden random variable?}
\label{Hidden variable}

In previous sections we presented a detailed study of the probability
distribution of MS packings in small 2d systems of bidisperse
frictionless disks.  We showed that the probabilities of individual
packings may differ by orders of magnitudes, not only as function of
packing fraction, but also in individual narrow packing-fraction
intervals.  Moreover, we identified important features of the packing
probabilities $\stateFrequency_k$ that are only weakly affected by the
details of the packing-generation protocol.

Since our packing-generation algorithms do not target any specific
packings, there must exist important properties of the
multidimensional configuration space that give rise to such widely
varying packing probabilities.  We have examined several local
properties of the PEL near the MS packing configurations and have
found gross correlations between these properties and the packing
probabilities.  However, there is large statistical scatter in the
data, and thus we conclude that none of the local features of PEL that
we have examined can fully explain the packing probability
distribution.

Yet, in spite of the complexity of the problem we have found several
interesting regularities.  In particular we have determined that the
probabilities $\stateFrequency_k$ in a narrow packing-fraction
interval, evaluated for different values of $N$, can be rescaled onto
a single master curve with the characteristic shape shown in Fig.\
\ref{sorted probabilities for different N} {(b)}.  In this section we
further explore this striking similarity of the sorted probability
distributions.  We base our analysis on a simple phenomenological
model that has been inspired, in part, by the correlation between the
packing probabilities and the minimal energy barriers [cf.\ Fig.\
\ref{barrier hight CG and MD} (b)].

In this model we characterize each MS packing by a set of $M$
independent continuous random variables
\begin{equation}
\label{random variables}
x_1,\ldots,x_M>0,
\end{equation}
all with the same probability distribution $\piOne(x)$, which is
approximately uniform for $x<x_{\rm max}$ and quickly decays beyond
$x_{\rm max}$.  The number of random variables $x_k$ is comparable
to the number of degrees of freedom in the system $M=O(Nd)$.
Our central assumption is that the probability $\stateFrequency_k$ of
a MS packing within the packing-fraction interval of interest is
controlled by the {\it smallest} of the random variables \ref{random
variables}:
\begin{equation}
\label{main assumption of model}
\stateFrequency_k\sim\min(x_1,\ldots,x_M).
\end{equation}

Using this assumption, the probability density $\piMin$ for
the random variable
\begin{equation}
\label{minimal x}
\xMin=\min(x_1,\ldots,x_M)
\end{equation}
can be written as
\begin{eqnarray}
\label{probability for minimal x}
\piMin(\xMin)&=&M\piOne(\xMin)\PiOne^{M-1}(\xMin)
\nonumber\\
&=&-\frac{d}{d\xMin}\PiOne^M(\xMin),
\end{eqnarray}
where
\begin{equation}
\label{probability  x not in interval}
\PiOne(y)=\int_y^\infty\piOne(x)dx
\end{equation}
is the probability that $x>y$.

From this result we can evaluate the expected value of the number of
MS packings $k(\bar y)$ that have $0<y<\bar y$,
\begin{equation}
\label{cumulative number of states in the model}
\frac{k(\bar y)}{k_{\rm max}}=\int_0^{\bar y}\piMin(\xMin)d\xMin
  =1-\PiOne^M(\bar y),
\end{equation}
where $k_{\rm max}$ is a total number of MS packings.

Now we apply the above results to the packing fraction interval in
which we have $k_{\rm max}=n_s(\Delta\phi)$ states.  After inserting
assumption \ref{main assumption of model} into \ref{cumulative
number of states in the model}, we obtain the expression 
\begin{equation}
\label{number of states for given f}
\frac{k}{k_{\rm max}}=1-\PiOne^M(a^{-1}\stateFrequency_k),
\end{equation}
where $a$ is the proportionality constant in Eq.\ \ref{main
assumption of model}.  Accordingly, for a given probability
distribution $\PiOne$, our analysis yields the relation between the
sorted probabilities $\stateFrequency_k$ and the index $k$ in the
sorted sequence of states.  This relation corresponds to the plot
shown in Fig.\ \ref{sorted probabilities for different N} (strictly
speaking, to its inverse).

Our theory cannot be directly verified without specifying the
distribution of the random variables $x_k$, and we do not have any
{\it a priori\/} information regarding this distribution.  Our goal
here, therefore, is more limited.  We simply want to determine whether
or not the numerical results shown in Fig.\ \ref{sorted probabilities
for different N} are consistent with our assumption that $\piOne(x)$
is a relatively uniform function in some range of $x$, outside of
which it quickly decays.

In making this assumption, we have in mind features of the PEL such as
the distance from a given MS packing to the passes in the rim of the
local potential-energy basin.  In each direction, the distance to the
rim is smaller than the particle diameter; it is also conceivable that
the distance to the closest pass determines the probability
$\stateFrequency_k$.  However, the specific meaning of the
hypothetical random variables \ref{random variables} in our
phenomenological theory at this point has not been fleshed out.

\begin{figure}
\scalebox{0.5}{\includegraphics{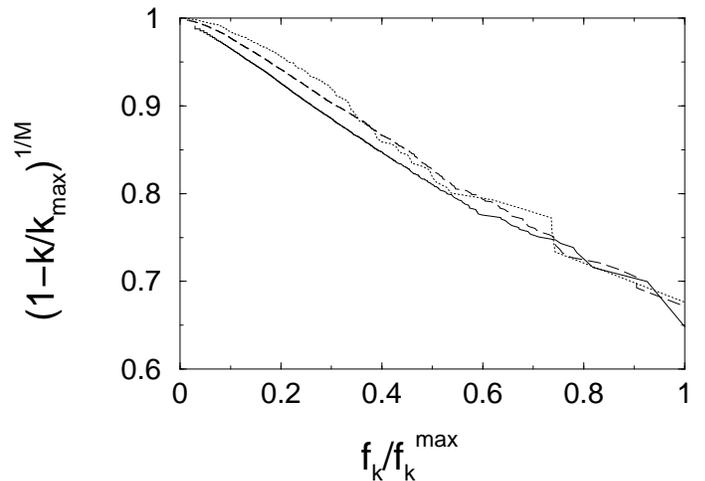}}
\caption{\label{speculation}
The right-hand side of Eq.~\ref{probability of x in terms of i}
$(1-k/k_{\rm max})^{1/M}$, with $M=dN-d$, is plotted versus the
normalized discrete packing frequencies
$\stateFrequency_k/\stateFrequency_k^{\rm max}$ over the narrow
packing fraction intervals in Fig.~\ref{sorted probabilities for
different N} using MS packings from the CG method in $N=10$ (dotted
line), $12$ (long-dashed line), and $14$ (solid line) particle
systems.
}
\end{figure}

To determine the approximate form of the probability distribution
$\PiOne$ from our numerical results, relation \ref{number of states
for given f} is inverted,
\begin{equation}
\label{probability of x in terms of i}
\PiOne(a^{-1}\stateFrequency_k)=\left(1-k/k_{\rm max}\right)^{1/M},
\end{equation}
and the data sets represented in Fig.\ \ref{sorted probabilities for
different N} are replotted in a form that emphasizes the structure
of our model.  Specifically, in Fig.~\ref{speculation}, the quantity
on the right-hand side of Eq.\ \ref{probability of x in terms of i}
is shown versus the normalized probability
$\stateFrequency_k/\stateFrequency_k^{\rm max}$ \cite{approx}.
The results in Fig.\ \ref{speculation} indicate that the transformed
quantity \ref{probability of x in terms of i} is a nearly linear
function of $\stateFrequency_k$.  By Eqs.\ \ref{probability x not in
interval} and \ref{probability of x in terms of i}, this linear
behavior is consistent with
\begin{equation}
\label{pi is theta}
\piOne(x)\sim\Theta(x),
\end{equation}
where $\Theta(x)$ is the Heaviside step function.  The form of the
probability distribution $\piOne$ determined from our numerical
results is thus compatible with assumptions of our model.

The results in Fig.\ \ref{speculation} cannot be treated as a direct
verification of our model, and the model itself is rather {\it ad
hoc\/}.  Nevertheless, the simplicity of the result \ref{pi is
theta} is notable.  We thus believe that our model captures at least
some essential aspects of the problem. However, the exact source of
the very broad distribution of the probabilities $\stateFrequency_k$
and the nature of the self-similarity of this distribution for
different values of $N$ (as revealed by the results shown in Fig.\
\ref{sorted probabilities for different N}) still remains an important
open problem.

\section{Conclusions and Future Directions}
\label{conclusions}

We have performed extensive numerical simulations with the aim of
generating mechanically stable (MS) packings of frictionless disks in
small bidisperse systems.  The MS packings are created using a
protocol in which we successively grow and shrink soft, purely
repulsive particles.  Each compression or decompression step is
followed by potential energy minimization, until all particle overlaps
are infinitesimal.  We focus on small systems with at most $14$
particles in 2D because in these systems we are able to find nearly
all distinct MS packings and can therefore accurately measure the
frequency with which each packing occurs.

One of the principal results in this work is that MS packing
frequencies differ by many orders of magnitude both as a function of
packing fraction and within narrow packing-fraction intervals.  We
have implemented here a fairly generic algorithm for generating MS
packings, and this algorithm does not specifically target
any of them.  Yet we find that packing frequencies are extremely
varied; moreover, the frequency variation increases with system
size.  We also find that the probability distribution can be scaled
onto a single master curve.

We argue that these striking results are important in a broader
context of theories of jammed granular media and glassy materials.  In
thermodynamic theories for dense granular media \cite{mehta} it is
usually assumed that MS packings within a small packing-fraction
interval are equally probable.  For our packing-generation protocol,
this assumption is certainly not valid.  Although we do not yet have
sufficient data, we believe that MS packings will also not be equally
probable for other commonly used experimental protocols, e.g.,
slow shear \cite{majmudar} or vibration under gravity \cite{nagel}.
However, thermodynamic theories based on the equal-probability
assumption are often applied to understand the properties of sheared
or vibrated granular materials \cite{makse2,barrat}.

Our findings show that the often-used equal-probability assumption for
stable grain packings in granular matter and inherent structures in
glass-forming liquids should be re-examined.  If it turns out that the
assumption is generally violated, except for some unphysical, highly
specialized algorithms (and we expect that this is the case),
thermodynamic theories of disordered granular packings will need to be
significantly reformulated.

In this work we focused entirely on a system of frictionless
particles.  However, we would like to point out that for frictional
particles there is another conceptual difficulty with the assumption of
equal-probability packings.  Since static friction can arrest
particle motion at different contact angles (analogous to a block
that can stop at any position on a wedge) MS packings of frictional
particles do not form points in configuration space, but rather
continuous hyper-surfaces.  Since packings of frictional particles are
often hyperstatic \cite{force}, the dimensionality of these
hyper-surfaces changes from packing to packing, and it is thus difficult
to introduce an appropriate probability measure.
   
Returning to the summary of the key results of our study, we note that
important features of the MS packing probabilities do not change
when we alter the packing-generation protocol.  Thus, we argue that
{\it protocol-independent} properties of configuration space must play
an important role in determining the MS packing frequencies.  To
investigate the connection between geometric properties of
configuration space and MS packing probabilities, we added thermal
energy to a set of MS packings and then measured several quantities as
the system fluctuated.  We monitored the time that elapsed before a MS
packing broke to a distinct one, which allowed us to determine
the heights of energy barriers that separate one MS packing from
another.  We also studied the displacement fluctuations in all
possible directions away from the original MS packing to infer crucial
features of the shape of the basin near each packing.

We found a gross correlation between the frequencies
$\stateFrequency_k$ of MS packings and the height of the lowest energy
barrier $\epsilon_0$ separating this packing from other ones.  The MD
packing frequencies roughly scale as $\stateFrequency_k \sim
\epsilon_0^{\lambda}$ with $\lambda \approx 2$, but there is
significant scatter in the data.  In addition, we found that the
eigenvalues of the displacement matrix scale as $d_i \sim
\epsilon_0^{\gamma_i}$ with $0.6 \alt \gamma_i \alt 1$.  These results
suggest that the MS packing frequencies are determined by one or a few
degrees of freedom, not by the local volume of configuration space
near each basin (in which case $\stateFrequency_k$ would scale much
more strongly with $\epsilon_0$).  However, the scatter in our data
implies that there are important unknown variables that are linked to
the MS packing probabilities.

Our results clearly indicate that this is complex problem and much
more work needs to be done to fully understand what determines the MS
packing frequencies.  Here we mention briefly some directions that we
are actively pursuing to address this question. 1) We are measuring
the hypervolumes of the regions in configuration space whose vertexes
are the nearby low-order saddle points of each MS packing.  We will
investigate the relation between these hypervolumes and packing
probabilities. 2) We are studying the ($dN-d$)-dimensional breaking
vector that connects the initial MS packing to the MS packing to which
it breaks.  We want to determine whether or not the breaking vector is
correlated with the directions of large displacements when the system
is thermally fluctuating.

\section*{Acknowledgments}

Financial support from NSF grant numbers CTS-0348175 (GG,JB),
DMR-0448838 (GG,CSO), and CTS-0456703 (CSO) is gratefully
acknowledged.  We thank N. Xu for providing us with some of the MS
packings obtained using the CG energy minimization method and
B. Chakraborty and N. Menon for helpful discussions. We also thank
Yale's High Performance Computing Center for generous amounts of
computer time.


\begin{references}
\bibitem{rare} We note that for $N=10$ there are a small number of
rare MS packings that have been generated by the MD but not by the CG
procedure.
\bibitem{boltz}
We have set Boltzmann's constant $k_b=1$ throughout.
\bibitem{approx}
The index $k$ in the sorted probability data sets are
discrete, and the probabilities $\stateFrequency_k$ correspond to a specific
realization of a random variables \ref{random variables}.  In
contrast, the corresponding quantities in Eqs.\ \ref{cumulative
number of states in the model}--\ref{probability of x in terms of i}
are average, continuous quantities.  To approximately compensate for
this discrepancy, we set $k_{\rm max}=n_s(\Delta\phi)+\delta$, where
$\delta\lesssim1\ll k_{\rm max}$ is a fitting parameter that is used
to adjust the position of the last few points on the high-probability
end of the plots.
\end{references}
\end{document}